\documentclass[11pt]{article}
\usepackage{float}
\usepackage{graphicx,graphics}
\usepackage{longtable}
\usepackage{titling}
\usepackage[margin=2cm,top=1.8cm,bottom=2.5cm,centering]{geometry}

\usepackage[utf8]{inputenc} 
\usepackage[T1]{fontenc} 
\usepackage[british]{babel} 
\usepackage{subfigure}
\usepackage[final]{pdfpages} 
\usepackage{pstricks}

\usepackage[colorlinks=true,urlcolor=black,linkcolor=black]{hyperref} 

\usepackage{lmodern} 

\usepackage{amsmath} 
\usepackage{esint} 
\usepackage{amssymb} 
\usepackage{esvect} 
\usepackage{tensor}
\usepackage{mathrsfs} 

\usepackage{array}
\usepackage{cellspace} 
\usepackage{makecell}
\usepackage{multirow}

\usepackage{tikz}
\usetikzlibrary{patterns}
\usetikzlibrary{calc,arrows,positioning}
\usetikzlibrary{arrows,shadows}
\usetikzlibrary{decorations.pathmorphing}	
\usetikzlibrary{decorations.markings}

\usepackage{textcomp}
\usepackage{graphicx}
\usepackage{url}
\usepackage{appendix}
\usepackage{subfigure}
\usepackage{pgfplots}
\pgfplotsset{compat=newest}
\usepackage{verbatim} 

\usepgfplotslibrary{patchplots}
\pgfplotsset{compat=1.3}
\numberwithin{equation}{section}

\usepackage{epstopdf}
\usepackage{listings}
\usepackage{lipsum}
\usepackage{colortbl}
\usepackage{qtree}
\usepackage{color}
\usepackage{gensymb}
\usepackage{enumerate}
\usepackage{numprint}
\usepackage{caption}
\usepackage[nottoc, notlof, notlot]{tocbibind} 
\usepackage{rotating} 

\newcommand{\R}{\mathbb{R}}

\newcommand{\Z}{\mathbb{Z}}

\usepackage{hhline}

\usepackage[final]{pdfpages}
\usepackage{titlesec}
\usepackage{stmaryrd}
\usepackage{delarray}

\newcommand{\dif}{\mathrm{d}}

\renewcommand{\i}{\mathrm{i}}
\newcommand{\pa}{\partial}
\renewcommand{\ge}{\geqslant}
\renewcommand{\le}{\leqslant}
\def\be{\begin{equation}}
\def\ee{\end{equation}}

\newcommand\bea{\begin{eqnarray}}
\newcommand\eea{\end{eqnarray}}
\newcommand\egal{&\!\!\!=\!\!\!&}
\def\cqfd{\hfill \hskip 1truemm \vrule height2.3mm depth0mm width2.3mm }

\begin{document}     

\title{\textbf{Concavity analysis of the tangent method}}

\date{}
\maketitle

\begin{center}
{\vspace{-19mm}\large \textsc{Bryan Debin}$^1$,\, \textsc{Etienne Granet}$^{2,3}$ \textsc{and Philippe Ruelle}$^1$}
\\[.8cm]
{\em {}$^1$Institut de Recherche en Math\'ematique et Physique\\ 
Universit\'e catholique de Louvain, Louvain-la-Neuve, B-1348, Belgium}
\\[.2cm]
{\em {}$^2$Institut de Physique Th\'eorique, Paris Saclay, CEA, CNRS,\\
F-91191 Gif-sur-Yvette, France}\\[0.2cm]
{\em {}$^3$Laboratoire de Physique Th\'eorique, D\'epartement de Physique de l'ENS,\\ 
\'Ecole Normale Sup\'erieure, Sorbonne Universit\'e, CNRS, PSL Research University, F-75005 Paris, France}
\\[.4cm] 
%
\end{center}

\vspace{1cm} 

\begin{abstract}
The tangent method has recently been devised by Colomo and Sportiello \cite{CS16} as an efficient way to determine the shape of arctic curves. Largely conjectural, it has been tested successfully in a variety of models. However no proof and no general geometric insight have been given so far, either to show its validity or to allow for an understanding of why the method actually works. In this paper, we propose a universal framework which accounts for the tangency part of the tangent method, whenever a formulation in terms of directed lattice paths is available. Our analysis shows that the key factor responsible for the tangency property is the concavity of the entropy (also called the Lagrangean function) of long random lattice paths. We extend the proof of the tangency to $q$-deformed paths.
\end{abstract}


\section{Introduction}

The phenomenon of phase separation and the emergence of limit shapes has been pointed out for the first time in the study of random domino tilings of Aztec diamonds \cite{CEP96,JPS98}. In this system, the random tilings exhibit two distinct phases, one external frozen (or solid) phase where the dominos are deterministically arranged as a brickwork, and one inner entropic, temperate (or liquid) phase with a non-zero density of horizontal dominos. The two phases are spatially separated by an interface which becomes sharper and sharper in the scaling limit, and converges to a circle, called the arctic circle. Since then, arctic phenomena generally refer to phase separations and arctic curves to lines separating distinct phases.

Arctic phenomena have been observed in many other models, including lozenge tilings of hexagons \cite{CLP98}, the 6-vertex model with domain wall boundary conditions \cite{CP10,CPZ10}, alternating sign matrices \cite{CP10b}, lozenge tilings of polygonal domains \cite{Pe14,DM15} and the 2-periodic Aztec diamond \cite{DK17}. For dimer models on bipartite planar graphs, rather general results have been obtained in \cite{KOS06,KO07}. The results obtained in these various models are technically involved and usually based on the calculation of appropriate bulk observables; in addition the methods used are specific to each model. 

Recently a general strategy has been proposed by Colomo and Sportiello \cite{CS16}. The tangent method, as it is known, was the result of a surprising geometric observation: these authors noticed that the arctic curve computed in \cite{CP10} for the 6-vertex model is the caustic of a family of straight lines determined by a one-point boundary observable. It was then elevated to a universal geometric principle and successfully checked in a number of situations where it indeed reproduced known results \cite{CS16,Ag18,DFL18,DFG18,DR18,DFG19}. It has been subsequently used to make predictions in cases the analytic shape of the arctic curve was not known, see \cite{CS16,DFL18,DFG19,CPS19,DFG19b,CKM19}.

The tangent method is as yet a heuristic but efficient method to obtain the analytic shape of arctic curves and is based on a few unproved though reasonable assumptions. The method itself as well as the underlying assumptions will be recalled in detail in Section \ref{sec5}. Let us simply say for now that it describes a way to produce a one-parameter family of curves (often straight lines), all tangent to the desired arctic curve, which can then be recovered as the envelope of that family. The concrete implementation of the method is model dependent but its principle is believed to be fairly universal. A minimal requirement is that the model under consideration can be formulated in terms of non-crossing (i.e. non-intersecting or osculating) random lattice paths. Apart from that, the domain of applicability is not precisely known.

A general, model independent proof of the assumptions which make the tangent method work is probably challengingly hard as it would be close to giving the shape of the arctic curve itself. Reference \cite{Ag18} presents a first rigorous (and rather technical) proof of these assumptions in a special model, namely the 6-vertex model at the ice point with particular boundary conditions. However the assumptions, namely that the arctic curve is deterministic in the scaling limit and is not affected by the displacement (or the addition) of a random path, can be separated from the conjecture stating the tangency property of the family of curves. The point of view which we propose here is that the tangency part of the tangent method may be addressed in a universal way, within the framework defined below.

Our approach may be summarized as follows. If one assumes that the arctic curve is deterministic and fixed, there remains only one random lattice path, travelling in a deterministic domain $\cal D$, bordered by pieces of the arctic curve and by the boundary in which the model is defined (the random path can be thought of as an extra path we introduce by hand). In known examples, this random path is made of elementary steps on a graph (usually $\Z^2$): in the Aztec diamond for instance, it is a Schr\"oder path with steps $(1,1)$, $(2,0)$ and $(1,-1)$, whereas for the 6-vertex model, it is a rotated Dyck path with possible steps  $(1,0)$ and $(0,1)$. As seen in the last two examples, the paths are directed, a condition we will assume throughout. 

The problem is then reduced to study, in the scaling limit, the probability distribution of a random path or random walk that stays inside the domain $\cal D$ (it cannot cross the arctic curve due to the non-crossing property of the paths underlying the arctic curve), the distribution being induced by the weights associated to the elementary steps, in finite number. While the endpoint is fixed and attached to the arctic curve, the starting point can be varied and taken away from the arctic curve. The distribution over random paths can be complicated on the lattice but dramatically simplifies in the scaling limit. By using a variational approach, we write the measure on continuous trajectories in terms of a classical action, and find that, in the scaling limit, this measure condensates on the unique trajectory which maximizes the action. We show that the unicity of the maximizing trajectory as well as its complete and explicit description follow from a central observation: the Lagrangean function defining the classical action is a strictly concave function, a property which we prove to hold for any directed random walk (on $\Z^2$ for simplicity). 

Applied to what the tangent method prescribes to do, these results imply that the scaling limit of the random path in $\cal D$ is, with probability 1, a straight line that hits the arctic curve tangentially and then coindices with it up to the (fixed) endpoint. Varying the starting point then yields a family of straight lines tangent to the arctic curve. We extend these results to the case where each random path receives an additional $q$-dependent weight taking into account the area below it, the so-called $q$-deformed case, considered in \cite{CY14,MP17,DFG19,DFG19b}. The main conclusion remains, namely after an appropriate rescaling of $q$, the random path in $\cal D$ collapses in the scaling limit onto a deterministic curve (it is no longer a straight line) which meets the arctic curve tangentially. 

The plan of the paper is as follows. Section 2 reviews the asymptotic combinatorics of directed random paths, introduces the entropy of paths with prescribed starting and ending points, and proves the main property underlying our analysis, namely that the entropy, soon to become a Lagrangean function, is a strictly concave function. In preparation for what has to come later, Section 3 computes passage probability distributions of random paths in the scaling limit. The variational approach, by which the measure of all the lattice paths which in the scaling limit collapse to a single continuous trajectory is written in terms of a classical action, is detailed in Section 4. Our main result, namely the tangency part of the tangent method, is presented in Section 5, while Section 6 contains a critical discussion of the method used to prove the tangency. The extension of the tangency property to the $q$-deformed case is carried out in Section 7.


\section{Partition functions for directed walks}
\label{sec2}

The general framework which we consider is that of lattice statistical models which can be formulated entirely in terms of non--crossing random paths, and showing the phenomenon of arctic curve. In addition, and this assumption is crucial, the lattice paths can be viewed as directed random walks in the plane, defined in terms of a set $S = \{\vec s_1 = (u_1,v_1),\, \vec s_2 = (u_2,v_2), \ldots, \vec s_k = (u_k,v_k)\}$ of elementary discrete steps and a set $W = \{w_1,\, w_2, \ldots, w_k\}$ of associated weights $w_i > 0$. We take $k\ge2$ as the case $k=1$ is trivial; for the same reason we also suppose that not all the $\vec s_i$ are collinear; likewise we may assume without loss of generality that one of the $w_i$ is equal to 1. For convenience we assume that the steps belong to $\Z^2$, although this should not be a strong restriction. 

By a directed walk, we mean a walk for which a return to a previously visited site is not possible (no erasure of any kind is performed). Thus the available steps are such that the condition $\sum_{i=1}^k a_i \vec s_i = 0$ with $a_i$ non--negative integers has the only solution $a_i=0$ for all $i$. Equivalently the convex hull of the points $\vec s_i$ in the plane does not contain the origin\footnote{The previous argument shows that the origin is not in the rational convex hull. It is not difficult to show that it can neither be in the real convex hull.}. This implies the existence of a straight line passing through the origin, which separates the plane in two halves, one of which strictly contains $S$, viewed as a set of points. Then a unit vector $\hat d \in \R^2$ exists such that all step vectors have scalar products with $\hat d$ which are strictly positive, $\vec s_i \cdot \hat d > 0$ for all $i$. For simplicity, we may assume that $S$ is such that the possible paths are contained in a cone which is itself contained in the right half-plane $x \ge 0$ (thus $u_i \ge 0$ for all $i$). We formulate here a number of preliminary results.

\subsection{The Lagrangean function of a walk} 

For the directed walk defined in terms of a set $S$ of possible steps $\vec s_i$ and weights $w_i$, we denote by $Z_{r,s}$ the weighted sum over all paths between the origin $(0,0)$ and the site $(r,s)$, the weight of a path being the product of the weights of its elementary steps. The assumptions made above imply $r \ge 0$. The generating function of the numbers $Z_{r,s}$ is given by 
\be
G(x,y) = \sum_{r,s} \: Z_{r,s} \, x^r \, y^s = \frac 1{1 - P(x,y)}, \qquad P(x,y) = \sum_{i=1}^k \: w_i \, x^{u_i} \, y^{v_i}, \qquad \vec s_i=(u_i,v_i).
\label{gen}
\ee
We will be mostly interested in the asymptotic behaviour of $Z_{r,s}$, when $r,s$ are both large with the ratio $s/r$ finite, and more specifically in its exponential growth rate.

By a standard argument, we note that for $m,n$ positive integers, the inequality $Z_{(m+n)a,(m+n)b} \ge Z_{ma,mb} \, Z_{na,nb}$ implies that the sequence $c_n = \log{Z_{na,nb}}$ is superadditive (satisfies $c_{m+n} \ge c_m + c_n$). By Fekete's lemma, the following limit exists 
\be
\lim_{n \to \infty} \frac1{n} \log Z_{na,nb} = \sup_n \frac1{n} \log Z_{na,nb} = L(a,b),
\ee
which we call the Lagrangean function for reasons that will be clear later on. The lemma does not guarantee that the limit is finite, but the following simple argument shows that it is finite, for all $a,b$, and grows at most linearly in $a$ and $b$.

Let $c = \min_{1 \le i \le k} (\vec s_i \cdot \hat d)$, a strictly positive number. To reach the point $(na,nb)$ requires at most $(na,nb)\cdot \hat d /c$ elementary steps, taken from the set $S$. We readily obtain the upper bound
\be
Z_{na,nb} \le \big(\sum_i w_i \big)^{(na,nb)\cdot \hat d /c} \quad \text{and} \quad L(a,b) \le \frac{(a,b)\cdot \hat d}{c} \log\big(\sum_i w_i\big).
\ee
Unless the points $(na,nb)$ are reachable for no $n$, the function $L(a,b) \ge 0$ is also bounded below by any non--zero term in the sequence $\frac1{n} \log Z_{na,nb}$.

Repeating the above arguments in terms $n'=na$ and final point $(n',n' \frac ba)$, we obtain the obvious but important scaling relation,
\be
L(a,b) = a \, L(1,\textstyle \frac ba) \equiv a L(\textstyle \frac ba).
\label{scaling}
\ee
In the rest of this article, the $L$ function will be used with a single argument, meant to be a slope. The support of the function of one variable $L(t)$ depends on the walk considered; it is given by the interval $[t_{\rm min},t_{\rm max}]$ where $t_{\rm min},t_{\rm max}$ are respectively the minimal and maximal slopes among the elementary available steps. Our assumptions imply that at least one of $t_{\rm min},t_{\rm max}$ is finite, so the support is closed or semi-closed.

In simple enough cases, the function $L(t)$ can be explicitly computed from the asymptotic form of the coefficients of the bivariate generating function $G(x,y)$, see for instance the review \cite{PW08} (see below for more details). As an illustration, we find the following Lagrangean functions for two simple walks. All weights $w_i$ are equal to 1 in these, except the horizontal step $(2,0)$ in the second example, which has weight $w$. For the second walk, we note that $Z_{na,nb}=0$ if $na+nb=1 \bmod 2$; in the other cases $L(a,b)=aL(\frac ba)$ with the function $L_2$ given below (provided $\frac ba \in [-1,1]$).

{\small

\be
\hspace{-1cm}
\psset{unit=5mm}
\begin{tikzpicture}[scale=0.5,baseline=4mm]
\draw[gray] (0,0)--(2,0);
\draw[gray] (0,1)--(2,1);
\draw[gray] (0,2)--(2,2);
\draw[gray] (0,0)--(0,2);
\draw[gray] (1,0)--(1,2);
\draw[gray] (2,0)--(2,2);
\draw[->,>=latex,thick,red] (0,1) -- (1,1);
\draw[->,>=latex,thick,red] (0,1)--(0,2);
\end{tikzpicture}
\quad P_1(x,y) = x + y\;:
\quad
L_1(t) = (1+t) \log{(1+t)} - t \log t, \qquad t \ge 0,
\label{ex1}
\ee

\bea
&&
\hspace{-1cm}
\psset{unit=5mm}
\begin{tikzpicture}[scale=0.5,baseline=4mm]
\draw[gray] (0,0)--(2,0);
\draw[gray] (0,1)--(2,1);
\draw[gray] (0,2)--(2,2);
\draw[gray] (0,0)--(0,2);
\draw[gray] (1,0)--(1,2);
\draw[gray] (2,0)--(2,2);
\draw[->,>=latex,thick,red] (0,1) -- (1,2);
\draw[->,>=latex,thick,red] (0,1)--(1,0);
\draw[->,>=latex,red,thick] (0,1)--(2,1) node[xshift=-0.2cm,below,black]{$w$};
\end{tikzpicture}
\quad P_2(x,y) = xy + x/y + wx^2\;:
\quad
L_2(t) = \log{\frac{\sqrt{1+w t^2}+\sqrt{w+1}}{\sqrt{1-t^2}}} + t \log{\frac{\sqrt{1+wt^2}-\sqrt{w+1}\,t}{\sqrt{1-t^2}}}, 
\nonumber\\
&& \hspace{12truecm} \qquad -1 \le t \le 1.
\label{ex2}
\eea
}

\subsection{Concavity} \label{twotwo}

For $(nx,ny)$ an intermediate passage point between $(0,0)$ and $(na,nb)$, the simple inequality (which can be made strict except in degenerate cases)
\be
Z_{na,nb} \ge Z_{nx,ny} \, Z_{n(a-x),n(b-y)},
\ee
immediately yields $L(a,b) \ge L(x,y) + L(a-x,b-y)$ (which cannot be made strict in general, as we will see later). Using the scaling relation (\ref{scaling}), it becomes
\be
L\left(\frac{b}{a}\right) \ge \frac{x}{a} L\left(\frac{y}{x}\right)+\left(1-\frac{x}{a}\right)L\left(\frac{b-y}{a-x}\right).
\ee
Setting $\alpha=\frac xa$, $t_1=\frac yx$ and $t_2=\frac{b-y}{a-x}$, the previous inequality is the statement that the function $L(t)$ is concave on its domain,
\be
L\big(\alpha t_1 + (1-\alpha) t_2\big) \ge \alpha L(t_1) + (1-\alpha) L(t_2), \qquad 0 \le \alpha \le 1.
\ee

However the function $L(t)$ satisfies a stronger property which turns out to be central in the subsequent arguments:  {\it the Lagrangean function $L(t)$ is strictly concave}. The strict concavity means that
\be
L\big(\alpha t_1 + (1-\alpha) t_2\big) > \alpha L(t_1) + (1-\alpha) L(t_2), \qquad 0 < \alpha < 1, \, t_1 \neq t_2.
\ee

The yet slightly stronger property, namely that $L''(t) < 0$ on its open support, which itself ensures the strict concavity, proves to be useful later on and can be obtained from the way the function $L(t)$ is to be computed. The reference \cite{PW08} reviews a number of methods to compute the asymptotic value of the coefficients of multivariate generating functions. In our case, the relevant function is the two-variable generating function given in (\ref{gen}), for which the results quoted in \cite{PW08} apply\footnote{In the survey \cite{PW08}, the generating function is assumed to be non singular at the origin and to expand in positive powers of $x$ and $y$ ($r,s \ge 0$). This assumption is not verified here if some of the elementary steps have a second component which is negative. When this is the case and since the step vectors are contained in a cone of aperture strictly less than $\pi$, they can be mapped by an affine transformation to steps contained in the first quadrant for which the results of \cite{PW08} apply. However, one can show that this procedure and a naive application of the results of \cite{PW08} yield exactly the same results. For this to hold, the directedness of the walks considered here is crucial.}. 

\begin{figure}[t]
\begin{tikzpicture}
\begin{axis}[
axis lines=middle
,x label style = {at={(axis description cs: 1.05,-0.05)}}
,xlabel={$t$}
,ylabel={}
,samples=41
,grid
,thick
,legend style={at={(0.95,0.5)},anchor=south east}
,domain=0:5
]
\addplot[no marks,orange] {1/(1 + x)};
\addlegendentry{$x_1(t)$}
\addplot[no marks,green] {x/(1 + x)};
\addlegendentry{$y_1(t)$}
\addplot[no marks, blue] {(1 + x)* ln(1 + x) - x *ln(x)};
\addlegendentry{$L_1(t)$}
\end{axis}

\begin{scope}[xshift=9.7cm]
\begin{axis}[
axis lines=middle
,x label style = {at={(axis description cs: 1,-0.1)}}
,xlabel={$t$}
,ylabel={}
,samples=41
,grid
,thick
,legend style={at={(0.28,0.7)},anchor=south east}
,domain=-1.01:1.01
,ymax=2
]
\addplot[no marks,orange] {sqrt((1 + 1)/(1 - x^2)) - sqrt((1 + x^2)/(1 - x^2))};
\addlegendentry{$x_2(t)$}
\addplot[no marks,green] {(sqrt(1 + x^2) + sqrt(1 + 1)* x)/sqrt(1 - x^2))};
\addlegendentry{$y_2(t)$}
\addplot[no marks, blue] {-ln(sqrt(1/(1 - x^2))*(sqrt(2) - sqrt(x^2 + 1))) -  x*ln((sqrt(x^2 + 1) + sqrt(2)*x)/sqrt(1 - x^2))};
\addlegendentry{$L_2(t)$}
\end{axis}
\end{scope}
\end{tikzpicture}
\caption{Plots of the functions $x(t)$, $y(t)$ and $L(t)$, for the two examples quoted in Section 2, first example on the left, second one, with $w=1$, on the right. The plots show the typical behaviour of these functions for semi-infinite and finite domains respectively.}
\label{fig1}
\end{figure}
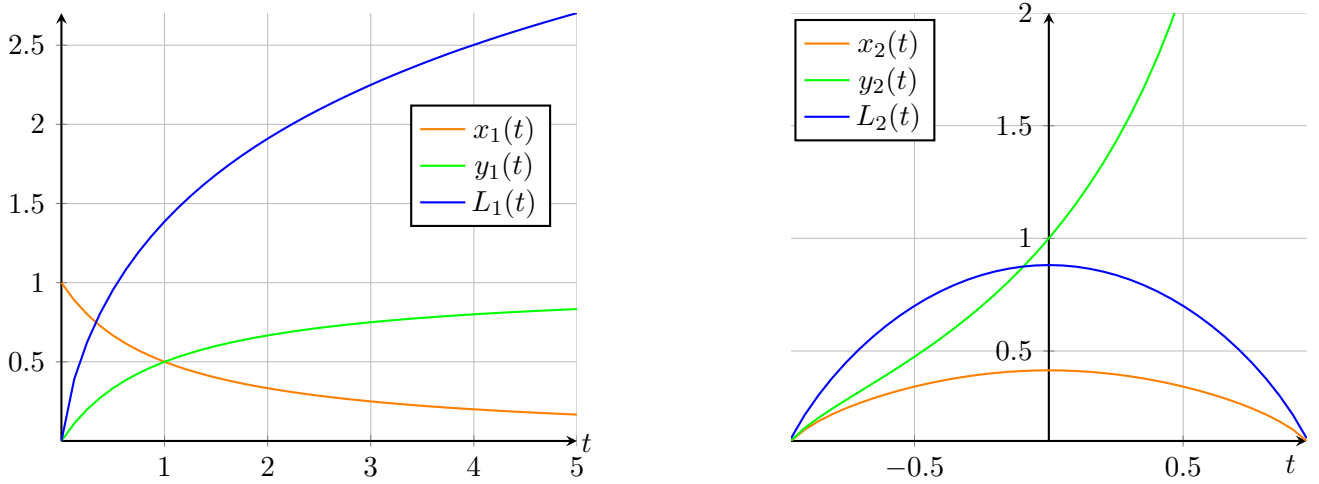

Let $x,y$ be two functions of $t$ satisfying the system (the indices of $P(x,y)$ denote partial derivatives)
\be
P(x,y) = 1, \qquad \qquad tx \, P_x(x,y) = y \,  P_y(x,y).
\label{syst}
\ee
Theorem 1.3 and Corollary 3.21 of \cite{PW08} state that (i) there is a unique solution for which $x(t),y(t)$ are real positive functions, and (ii) the asymptotic value of $Z_{na,nb}$, if non--zero, is given by
\be
Z_{na,nb} = \sqrt{\frac {N(t)}{2\pi n a}} \, \exp{\!\big\{na L(t)\big\}}, \qquad t = \frac ba.
\label{asympt}
\ee
Here $N(t)$ is an explicit function of $x(t)$ and $y(t)$, generally a rather complicated algebraic function of $t$, and 
\be
L(t) = -\log x(t) - t\, \log y(t).
\ee
In the two examples mentioned in the previous subsection, these functions are found to be equal to
{\small
\bea
&& \hspace{-.5cm} x_1(t) = \frac 1{1+t}, \quad y_1(t) = \frac t{1+t}, \quad N_1(t) = \frac{1+t}t, \qquad t > 0,\\
\noalign{\medskip}
&& \hspace{-.5cm} x_2(t) = \frac{\sqrt{1+w} - \sqrt{1+wt^2}}{w\sqrt{1-t^2}}, \quad y_2(t) = \frac{t\sqrt{1+w} + \sqrt{1+wt^2}}{\sqrt{1-t^2}}, \quad N_2(t) = \frac{(\sqrt{w+1}+\sqrt{1+wt^2})^2}{\sqrt{w+1} \,\sqrt{1+wt^2}\, (1-t^2)}, \quad -1 < t < 1. \nonumber \\
\eea
}
The corresponding functions $x(t), y(t), L(t)$ are shown in Figure \ref{fig1}.

Before proving that $L''(t)$ is strictly negative, we show that the three functions $x(t), y(t), L(t)$ are smooth. The first equation of (\ref{syst}) yields a first implicit equation $f_1(x,y)=0$, where $\partial_x f_1(x,y)$ nowhere vanishes in the domain $x,y > 0$. Moreover since $f_1(x,y)$ is $C^\infty$ in both variables in the same domain, the implicit function theorem implies that $x=x(y)$ is a $C^\infty$ function of $y$. Similarly the second equation in (\ref{syst}), after substituting $x=x(y)$, yields an implicit equation $f_2(y,t)=0$, which is $C^\infty$ in both $t$ and $y>0$. The derivative $\partial_y f_2(y,t)$ is found to be equal to 
\be
\partial_y f_2(y,t) = - \frac{t^2x}y \big[ P_x(x,y) + x P_{xx}(x,y) \big] + 2 t x P_{xy}(x,y) - P_y(x,y) - y P_{yy}(x,y),
\label{df2}
\ee
where we used the identity 
\be
x'(y) = - \frac {tx}y,
\label{xprime}
\ee
obtained by differentiating the first equation of (\ref{syst}) with respect to $y$, and using the second equation (here and in what follows, $f'$ denotes the derivative of $f$ with respect to its argument, made explicit). Recalling the form of the function 
\be
P(x,y) = \sum_{i=1}^k \: w_i \, x^{u_i} \, y^{v_i}, \qquad u_i \ge 0,
\ee
a simple computation shows that 
\be
- \frac{t^2x}y \big[ P_x(x,y) + x P_{xx}(x,y) \big] + 2 t x P_{xy}(x,y) - P_y(x,y) - y P_{yy}(x,y) = -\sum_{i=1}^k \: w_i \, (v_i - t u_i)^2 \, x^{u_i} \, y^{v_i-1}
\ee
is strictly negative. Thus $\partial_y f_2(y,t)<0$ nowhere vanishes in the domain $y>0$, $t \in \R$. A second use of the implicit function theorem implies that $y=y(t)$ is a smooth function of $t$ on its open domain $]t_{\rm min},t_{\rm max}[$. The statement that the functions $x(t)$ and $L(t)$ are smooth readily follows. We note that the double use of the implicit function theorem readily proves the uniqueness of the real functions $x(t),y(t)>0$.

In order to check that $L''(t) < 0$, we differentiate the first equation of (\ref{syst}) with respect to $t$ (after the substitution $x=x(t)$ and $y=y(t)$) and use the second equation to get the condition $(\log x(t))' = -t \, (\log y(t))'$. It allows to write the first two derivatives of $L(t)$ as
\be
L'(t) = -\log y(t), \qquad L''(t) = -\frac{y'(t)}{y(t)}.
\ee
Differentiating now the second equation of (\ref{syst}) with respect to $t$ and using (\ref{xprime}) once more yields
\be
y'(t) = \frac{x P_{x}(x,y)}{P_y(x,y) + y P_{yy}(x,y) + \frac{t^2x}y \big[ P_x(x,y) + x P_{xx}(x,y) \big] - 2 t x P_{xy}(x,y)}.
\label{frac}
\ee
From the specific form of $P(x,y)$, we see that $xP_x(x,y)$ is strictly positive in the domain $x,y > 0$, whereas the denominator, equal to $-\partial_y f_2(y,t)$ computed above, is also strictly positive. Thus the numerator and denominator of (\ref{frac}) are both strictly positive for $x,y > 0$, which implies $y'(t) > 0$ and thus $L''(t) < 0$ as claimed. 


\section{Passage probability distributions}
\label{sec3}

In this section, we prove that in the scaling limit, by which all large lattice distances $na,nb,\ldots$ are rescaled to $a,b,\ldots$ (equivalently, the lattice spacing is shrunk from 1 to $\tfrac 1n$), the probability measure on paths connecting two points concentrates on a unique trajectory, namely a straight line. We also investigate the size of the fluctuations around the straight line.

\subsection{Passage probabilities}

Given a distant site $(na,nb)$, the probability that a directed walk passes through the point $(nx,ny)$ is given by
\bea
{\mathbb P}_0\big[(nx,ny)\big|(na,nb)\big] \egal \frac{Z_{nx,ny} \, Z_{n(a-x),n(b-y)}}{Z_{na,nb}}\nonumber\\
&\!\!\simeq\!\!&  \exp \left\{ a n\left[ \frac{x}{a} L\left(\frac{y}{x}\right)+\left(1-\frac{x}{a}\right)L\left(\frac{b-y}{a-x}\right)-  L\left(\frac{b}{a}\right)\right]\right\},
\label{F}
\eea
up to the multiplicative non--exponential terms given in (\ref{asympt}). The strict concavity of the Lagrangean function implies that the combination in the square brackets is strictly negative, except when the two points $\frac yx$ and $\frac{b-y}{a-x}$ coincide, namely when $\frac yx = \frac ba$, in which case it trivially vanishes. We therefore obtain that the above probability is exponentially depreciated with $n$ when the intermediate point is not on the straight line, and that it is asymptotic to 1 otherwise. An analogous though somewhat weaker result has been proved in \cite{DFL18}. 

\subsection{Fluctuations}
\label{sec32}

We have seen that in the rescaled domain, the path between $(0,0)$ and $(a,b)$ is almost surely a straight line. However at finite but large $n$, the distribution shows fluctuations. It is natural and important for what follows to understand the size of these fluctuations.

In the asymptotic regime, the passage probability is given by
\be
{\mathbb P}_0\big[(nx,ny)\big|(na,nb)\big] = \sqrt{\frac{a N(\frac yx) N(\frac{b-y}{a-x})}{2\pi n x(a-x) N(\frac ba)}} \times \exp{\{an F(y)\}},
\ee
where $F(y)$ is the combination of three terms in the square brackets in (\ref{F}). As noted above, $F(y)$ is strictly negative except at $y(x) = tx = \frac {bx}a$ where it vanishes. Because of the factor $n$ in the exponential, only the values of $y$ very close to $y(x)$ will make a significant contribution to the probability. Therefore we can expand, to leading order, $F(y)$ and the factor in front of the exponential around $y(x)$. From above we know that $F(tx) = 0$ but also that $F'(tx) = 0$ since $F(y)$ is strictly negative away from $y=tx$, which is therefore a maximum. Moreover a simple computation yields $F''(tx) = \frac 1{x(a-x)} L''(t) < 0$, a strictly negative number according to the results of the previous section. We therefore obtain the following Gaussian approximation,
\be
{\mathbb P}_0\big[(nx,ny)\big|(na,nb)\big] = \frac 1n \sqrt{\frac1{2\pi \sigma^2}} \sqrt{\frac{N(t)}{-L''(t)}} \exp{\left\{-\frac{(y-tx)^2}{2\sigma^2}\right\}},
\label{gauss}
\ee
for which we allow $y$ to take all real values (rather than those related to the reachability to and from $(nx,ny)$), and where
\be
\sigma = \sqrt{\frac{x(a-x)}{-anL''(t)}}.
\label{variance}
\ee

The above probability is not normalized upon summation over $y$, since one readily obtains
\be
\sum_{ny=-xt_{\rm min}}^{ny=-xt_{\rm max}} {\mathbb P}_0\big[(nx,ny)\big|(na,nb)\big] \simeq \int_{-\infty}^\infty {\rm d}(ny) \: {\mathbb P}_0\big[(nx,ny)\big|(na,nb)\big] = \sqrt{\frac{N(t)}{-L''(t)}}.
\ee
On general grounds, we do not expect it to be normalized. Indeed if a vertical step of the form $\vec s=(0,v)$ is allowed (has non--zero weight), the walk may visit several values of $y$ for the same value of $x$, causing an overcounting (integral larger than 1). Likewise, if some of the possible steps have horizontal components $u \ge 2$, a fraction of walks will not visit any site on the vertical line at $x$, causing this time an undercounting (integral smaller than 1). Finally, there may be arithmetical constraints on the sites $(x,y)$ that are actually reachable, in which case the summation over $y$ must be further restricted. These various situations are reflected by the presence of the factor $\sqrt{N(t)/(-L''(t))}$ in (\ref{gauss}). 

\begin{figure}[t]
\hspace{-.2cm}
\begin{tikzpicture}[scale=0.77]
\begin{axis}[xscale=1.9,yscale=1.2
,xlabel={$ny$}
,y label style = {at={(axis description cs: -0.035,0.25)},anchor=south west}
,ylabel={$\mathbb{P}[(nx,ny)|(n,n)]$}
,yticklabel style={/pgf/number format/fixed}
,ytick={0.05,0.10,0.15}
,xmin=0
,xmax=200,
,ymin=0
,legend style={at={(0.15,0.45)},anchor=south west,legend columns=2}]
\addlegendimage{empty legend}
\addlegendimage{empty legend}

\addplot[ybar,ybar legend,bar width=.3pt,blue,fill=blue,opacity=0.8] table {hist_x20_N200.txt};  
\addplot[no marks,blue!70!black] table {hist_x20_N200_th.txt};    

\addplot[ybar,ybar legend,bar width=.3pt,purple,fill=purple,opacity=0.5] table {hist_x50_N200.txt}; 
\addplot[no marks,purple!70!black] table {hist_x50_N200_th.txt};

\addplot[ybar,ybar legend,bar width=.3pt,red,fill=red,opacity=0.5] table {hist_x100_N200.txt};  
\addplot[no marks,red!70!black] table {hist_x100_N200_th.txt}; 
   
\addplot[ybar,ybar legend,bar width=.3pt,orange,fill=orange,opacity=0.5] table {hist_x150_N200.txt};
\addplot[no marks,orange!70!black] table {hist_x150_N200_th.txt};   
   
\addplot[ybar,ybar legend,bar width=.3pt,yellow,fill=yellow,opacity=0.5] table {hist_x180_N200.txt};   
\addplot[no marks,yellow!70!black] table {hist_x180_N200_th.txt};
   
\legend{simulation, theory,$nx=20$,$nx=20$,$nx=50$,$nx=50$,$nx=100$,$nx=100$,$nx=150$,$nx=150$,$nx=180$,$nx=180$}
\end{axis}
\begin{scope}[xshift=15cm,yscale=1.2]
\begin{axis}[
axis lines=middle
,xlabel={$x$}
,ylabel={$n \; \sigma^2(x)$}
,x label style = {at={(axis description cs: 0.55,-0.15)}}
,y label style = {at={(axis description cs: -0.1,0.4)},rotate=90,anchor=south west}
,xtick={0,0.2,0.4,0.6,0.8,1}
,grid
,thick
,legend style={at={(0.68,0.21)},anchor=south east}
,xmin=0,xmax=1
]
\addplot[only marks,mark=square,blue] table {variance_free_N25.txt};
\addlegendentry{$n=25$}
\addplot[only marks,mark=triangle,orange,mark repeat=2] table {variance_free_N50.txt};
\addlegendentry{$n=50$}
\addplot[only marks,mark=diamond,green,mark repeat=4] table {variance_free_N100.txt};
\addlegendentry{$n=100$}
\addplot[only marks,mark=star,red,mark repeat=8] table {variance_free_N200.txt};
\addlegendentry{$n=200$}
\addplot[domain=0:1] {2*x*(1-x)};
\addlegendentry{$2x(1-x)$}
\end{axis}
\end{scope}
\end{tikzpicture}
\caption{The left panel shows the probability distributions of a passage at $y$ for five different values of $x$ ($a=b=1$), for the directed paths defined by $S=\{(1,0),(0,1)\}$, each step having weight $w=1$ (example 1 in Section \ref{sec2}). The bars are the results of simulations (36\,000 samples) for $n=200$, whereas the solid lines represent the Gaussian approximation (\ref{gauss}). The right panel compares, for the same paths but for different values of $n$, the quantity $n\sigma^2$ obtained from simulations with the one computed in (\ref{variance}) ($a=b=1$, $t=1$ and from (\ref{ex1}), $L_1''(1)=-\tfrac12)$.}
\end{figure}
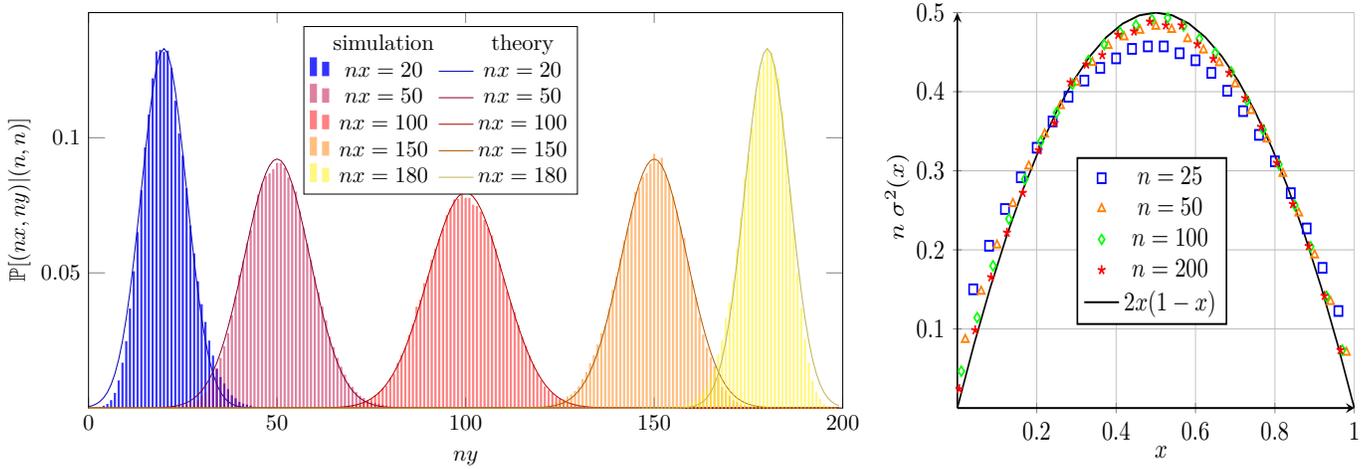

Despite the (expected) absence of a proper normalization, the distribution (\ref{gauss}) shows a dispersion of order $n^{-1/2}$ around the mean value $y=tx$, in the rescaled domain. On the lattice, keeping the large variables $nx,ny,na,nb$, we expect a standard deviation of order $n^{1/2}$. This is well confirmed by numerical simulations, as shown in Figure 2 for the lattice paths of example 1 given in (\ref{ex1}).


\section{Variational approach}
\label{sec4}

In this section, we would like to reformulate the problems and results developed in the previous sections as a continuous variational problem. Rather than counting the possible paths within a certain class dictated by the choice of possible elementary steps, and then doing statistical estimates about passage probabilities, we associate, in the continuum limit,   a classical action to each path. The action allows to assign a statistical weight to a whole trajectory, represented as a continuous function $h(x)$, the weight being itself inherited from those of the lattice elementary steps. The resulting variational problem appears to be the appropriate tool to determine the extremal paths used in the tangent method to characterize arctic curves. 

Let us consider all possible paths from (0,0) to $(na,nb)$, made of elementary steps taken from the set $S$. For $n$ large, their weighted sum $Z_{na,nb}$ has the asymptotic value given in (\ref{asympt}). Let us now focus on all the paths which pass through $K+1$ intermediate points $\vec x_j$, with $\vec x_j = (\frac {ja}K n,h_j n)$ for $j=0,\ldots,K$. Here the $h_j$ are fixed numbers independent of $n$ satisfying $h_0=0$ and $h_K=b$. Provided $\frac nK$ is large, the weighted sum $Z_n[\{h_j\}]$ over such paths is
\be
Z_n[\{h_j\}] = \prod_{j=0}^{K-1} \: Z_{(\frac {an}K,(h_{j+1}-h_j)n)} = \sqrt{\prod_{j=0}^{K-1} \, \frac {K N(t_j)}{2\pi n a}} \, \exp \Bigg\{ \frac{an}{K} \sum_{j=0}^{K-1} L(t_j) \Bigg\}, \qquad t_j = \frac{(h_{j+1}-h_j)K}a.
\label{eq_weighted_sum_q1}
\ee
Assuming that $h_j$ is the value at $\frac{ja}K$ of some real function\footnote{This is largely legitimate because the paths we consider are directed to the right. In case vertical steps are possible, a vertical macroscopic stretch is unlikely.} $h$, namely $h(\frac{ja}K)=h_j$, we obtain, for large enough $K$, that $t_j \simeq h'(\frac{ja}K)$, and that the sum over $j$ in the exponential is well approximated by an integral, leading to
\be
Z_n[h] = \sqrt{\prod_{j=0}^{K-1} \, \frac {K N(t_j)}{2\pi n a}} \times \exp \left(n \int_0^a  \dif x \:L\big(h'(x)\big) \right).
\ee
The function $h$ needs not be everywhere differentiable. Looking at the way it was introduced, the mere existence of a right derivative is sufficient. More generally, we may consider the set of functions $h$ which have a left and a right derivative, hence piecewise of class ${\mathcal C}^1$.

The other dependence of $Z_n[h]$ in the trajectory is in the factor in front of the exponential, which can similarly be written as
\be
\sqrt{\prod_{j=0}^{K-1} \, \frac {K N(t_j)}{2\pi n a}} \simeq \exp \left(\frac K{2a} \int_0^a  \dif x \: \log{\frac{K N\big(h'(x)\big)}{2\pi na}} \right).
\ee
Since $\frac Kn$ was assumed to be small (we will argue that choosing $K$ proportional to $\sqrt{n}$ is good enough), this term is subdominant, as expected.

We obtain that the fraction of paths going from the origin to the point $(na,nb)$, and which collapse in the scaling limit to a continuous trajectory described by the function $h(x)$, for $x \in [0,a]$, $h(0)=0$ and $h(a)=b$, has a relative statistical weight given in terms of an action $S[h]$ associated with $h$, 
\be
Z_n[h] \simeq e^{nS[h]}\,, \qquad S[h] = \int_0^a \dif x \:L\big(h'(x)\big).
\label{eq_non_deformed_action}
\ee
Interestingly the weights of the paths depend on the underlying lattice walk only through the function $L$ (which was called Lagrangean for obvious reasons). A trivial but important remark is that not every function $h$ is permitted: it must be compatible with the way the discrete walk has been defined, implying that $h'(x)$ must be in the interval $[t_{\rm min},t_{\rm max}]$ for all $x$.

Because the Lagrangean function depends on the derivative of $h$ only, the Euler-Lagrange equation, computed on the subdomains where $h$ is $\mathcal C^1$, takes the simple form
\be
L''(h') \, h''(x) = 0, \qquad h(0)=0, \, h(a)=b.
\ee
As $L''$ is strictly negative on its (open) domain, the trajectories $h(x)$ which make the action extremal satisfy $h''(x)=0$, that is, $h(x)$ is piecewise rectilinear. The concavity of $L$ ensures that the straight line $h(x) = bx/a$ is the unique global maximum of $S[h]$, as proved below. In the limit $n \to \infty$, the straight line is therefore observed almost surely, consistently with the results obtained in Section \ref{sec3}.


\section{The tangent method}
\label{sec5}

The applicability conditions of the tangent method are not precisely known, but it is generally expected to be valid in models which can be formulated in terms of random non-crossing lattice paths (or at least showing some form of repulsion). In these models, the arctic phenomenon arises when a large number of paths accumulate to form, in the scaling limit, a sharp and deterministic interface between a frozen region, which contains no path at all, and an entropic region, densely filled with paths. Focusing on a piece of the arctic curve connecting two points on the boundary of the domain, say $x_0$ and $x_1$, we would see, close to the scaling limit, a compact cluster of paths starting from a neighbourhood of $x_0$ and ending at sites in a neighbourhood of $x_1$. In particular the outermost path would start from site $b_0$ and go to site $b_1$, in the scaling neighbourhood of $x_0$ and $x_1$ respectively.

\begin{figure}[t]
\begin{center}
\begin{tikzpicture}[scale=0.7]
\begin{scope}
\draw (0,0)--++(0,5)--++(5,0);
\draw[blue,scale=1,domain=0:5,smooth,variable=\x] node[below]{$x_0$} plot ({\x},{5*sqrt(1-(\x-5)^2/25)}) node[below]{$x_1$};
\draw[red,scale=1,domain=0:40/29),smooth,variable=\x] plot ({\x},{2+0.05+21/20*(\x+0.05)}) ;
\draw[red,dashed,scale=1,domain=0:5-sqrt(199)/20,smooth,variable=\x] plot ({\x},{5*sqrt(1-(\x-5+0.05)^2/25)+0.05});
\draw[red,scale=1,domain=40/29:5-sqrt(199)/20,smooth,variable=\x]  plot ({\x},{5*sqrt(1-(\x-5+0.05)^2/25)+0.05});
\draw[red] (-1/2,3/4) node {$b_0$};
\draw[red] (-1/2,2) node {$b_0'$};
\draw[red] (4+1/4,5) node[above] {$b_1$};
\end{scope}
\end{tikzpicture}
\end{center}
\caption{Sketch of the arctic curve (blue), the boundaries of the domain (black), and the paths (red).}
\label{fig3av}
\end{figure}
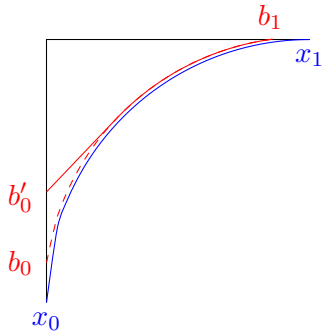

\begin{figure}[t]
\begin{center}
\begin{tikzpicture}
\clip (-0.5,-1) rectangle (25,-9);
\draw (0,-5) node{$X_1$};
\draw (3,-1.3) node{$X_2$};
\includegraphics[scale=0.35,angle=-90]{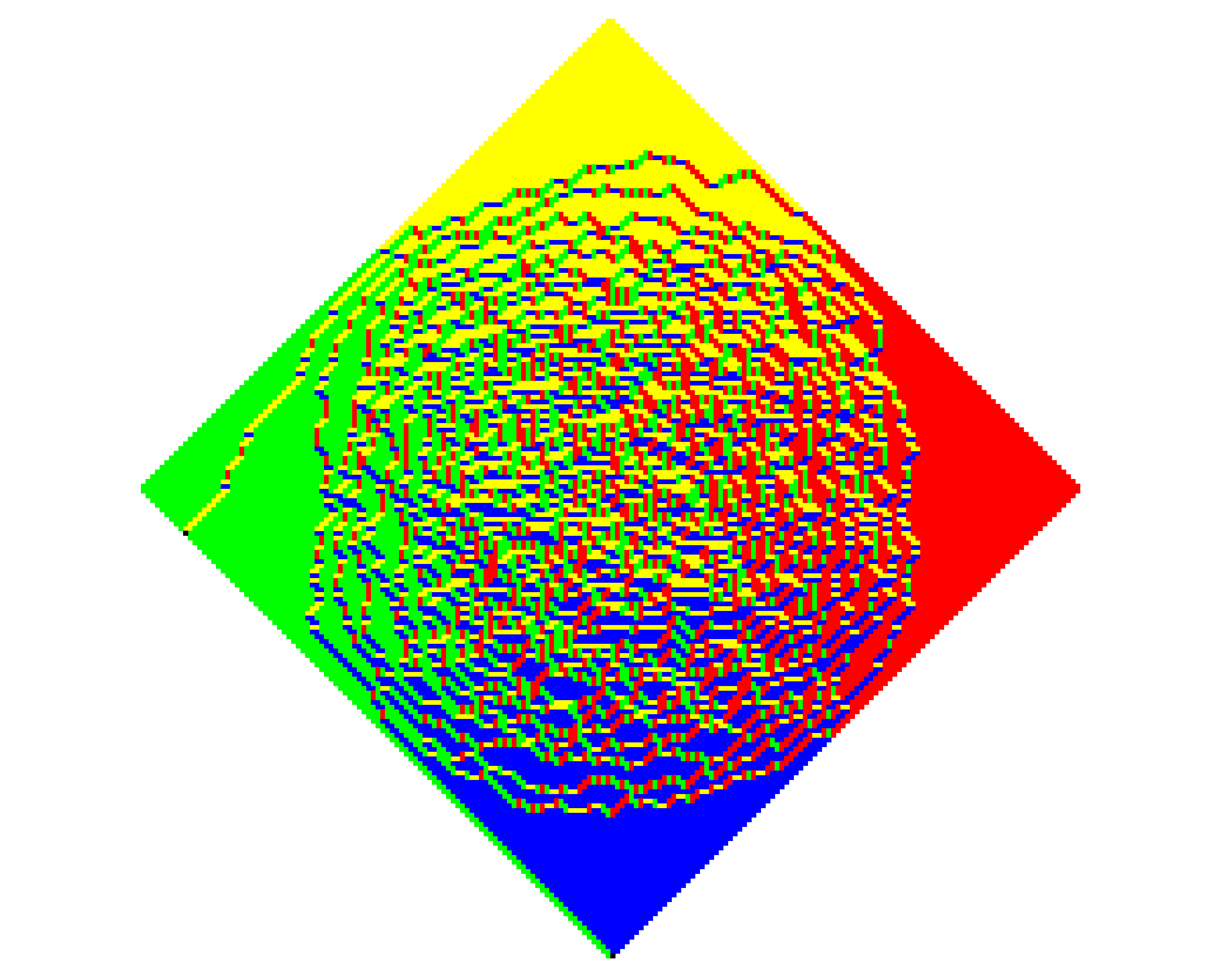};
\hspace*{2cm}\includegraphics[scale=0.35,angle=-90]{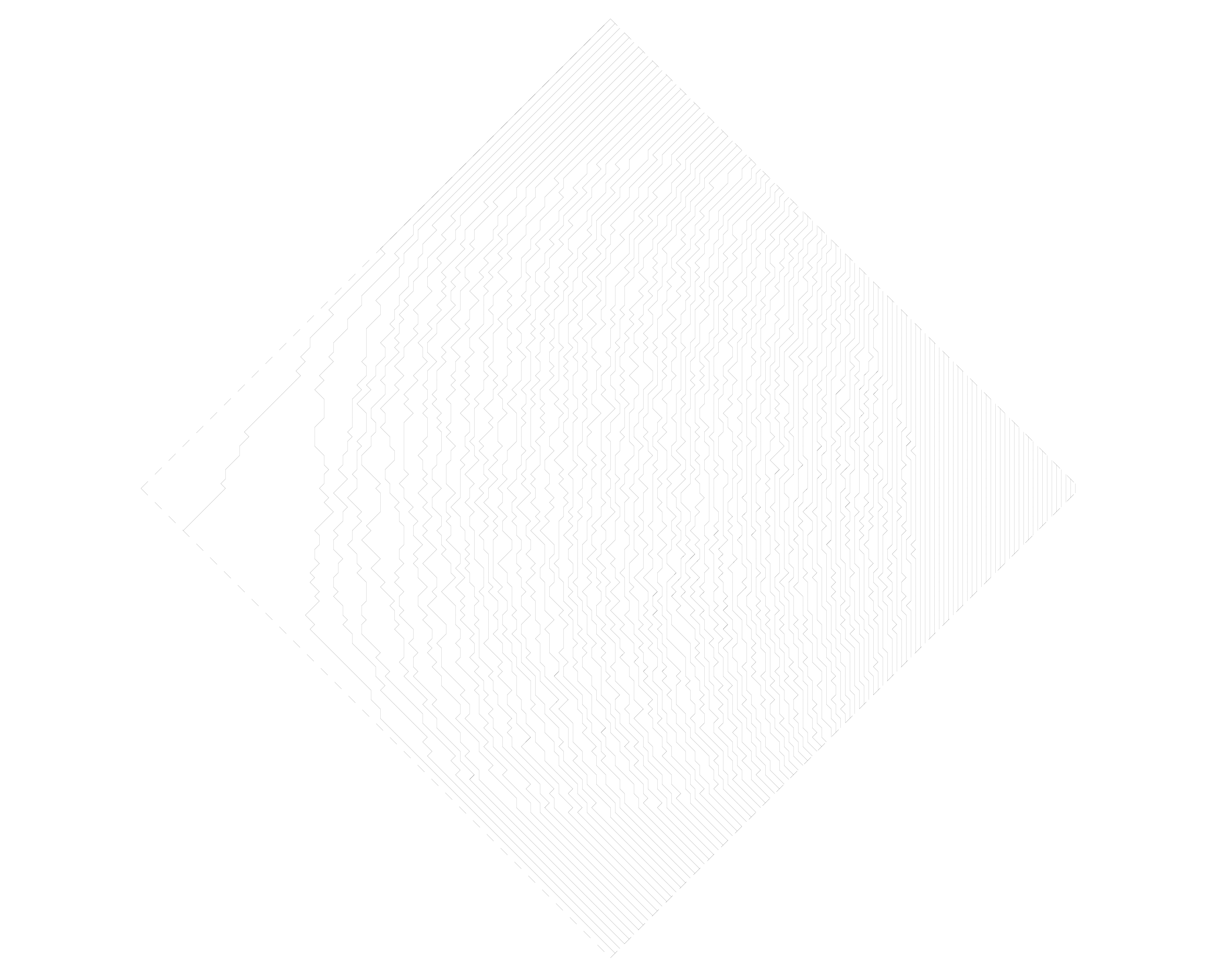};
\end{tikzpicture}
\end{center}
\vspace{-1truecm}
\caption{On an Aztec diamond of order 100, two monomers have been inserted on the upper left side to force the outermost path to start from a new location instead of the left corner. The new tiling configuration is shown on the left, the corresponding path configuration on the right. The two monomers at $X_1$ and $X_2$ are drawn in black. Before the change, the paths all start from the lower left side and end on the lower right side of the diamond.}
\label{fig3}
\end{figure}

The tangent method requires to slightly modify the model in such a way that the starting (or ending) point of the outermost path is moved from $b_0$ to another point $b'_0$, located on the boundary between $x_0$ and $x_1$ and at a macroscopic distance from $b_0$, see Figure \ref{fig3av}. The specific way this change is implemented depends on the model under consideration. In one of the best known cases, namely the Aztec diamond, one can achieve this by inserting two monomers, as shown in Figure \ref{fig3}. As a result of this change, the outermost path, now starting from $b'_0$ but still tied to its ending point $b_1$, will first traverse the frozen region, then approach and merge in the compact cluster of paths, and eventually reach its endpoint $b_1$ along with the rest of the cluster.

The tangent method relies on one assumption and one conjecture, both supposedly valid in the scaling limit.

\begin{itemize}
\item[(i)] It makes the working hypothesis that the change of starting point of the outermost path, from $b_0$ to $b'_0$, has so little influence on the other paths that the arctic curve itself is not affected. If this were not the case, the method would predict a different arctic curve, as compared to other methods.
\item[(ii)] It postulates that the outermost path is almost surely a straight line traversing the frozen region from $b'_0$ until it meets the arctic curve tangentially. After the meeting point, the path coincides with the arctic curve till the endpoint $x_1$. In fact, the same conjecture more generally holds if the starting point is taken out of the domain, namely to a point $a_0$ chosen in such a way that the paths can enter the domain in the frozen region (through a boundary point between $x_0$ and $x_1$). Let us observe that the tangency of the straight lines requires local curvature properties of the arctic curve (the tangent lines must lie in the frozen region).
\end{itemize}

The efficiency of the method relies on the postulate (ii). By varying the starting point $b'_0$, one obtains a one-parameter family of straight lines tangent to the arctic curve, which is then simply recovered as its envelope. We note that in some cases, like for instance in the model of $q$-weighted lattice paths studied in \cite{DFG19,DFG19b}, the straight lines are replaced by other known curves; however the method remains and the way the arctic curve is retrieved as the envelope of a family of tangent curves is the same.

Clearly, if the arctic curve is known, it is trivial to compute the family of tangential lines, each of which arrives at a given point $b'_0$ of the boundary. The tangent method aims at solving the inverse problem: given a point $b'_0$, find the slope of the line starting from $b'_0$ and hitting the arctic curve tangentially. This is usually done by resorting to a lattice point $a_0$ outside the domain, as mentioned in (ii), and by considering the paths from $a_0$ to $b_1$ which enter into the domain through a (random) boundary point between $x_0$ and $x_1$. In the scaling limit and given $a_0$, the paths are straight lines with probability 1, entering the domain at some boundary point and hit the arctic curve tangentially. Since the probability distribution over the point of entry concentrates on a unique point in the scaling limit, call it $b'_0$, it can be computed on the lattice as the entry point of maximal probability. In this way, the point $b'_0$ can be obtained as a function of $a_0$ and allows to compute the slope of the line that will eventually meet the arctic curve tangentially. Varying $a_0$ yields the required family of tangent lines.

As pointed out in \cite{CS16}, the hypothesis (i) is hard to formalize in general, let alone to prove. 
The conjecture (ii) is easier to formulate in general; it is our main purpose to provide strong arguments in favour of it.

Assuming the hypothesis (i), we view the compact cluster of paths collapsing to the arctic curve $\mathcal C$ in the scaling limit, as being given. It influences the shape and the distribution of the outermost path because it acts as an impenetrable frontier, due to the non-crossing property built in the model. The problem for the outermost path can then be posed in the following terms: {\it what is the distribution on the lattice paths from $a_0$ to $b_1$ (same notations as above), made of elementary steps from the set $S$, and constrained not to cross a certain curve $\mathcal C$ ? In particular, in the scaling limit, does this distribution condensate on a deterministic trajectory with the properties given in (ii) ?}

Compared to the results obtained in the previous sections, the novel issue is of course the presence of the impenetrable curve $\mathcal C$ because in actual situations, the endpoint $b_1$ cannot be reached from $a_0$ by following a straight line, as it would cross $\mathcal C$. As the arctic curve is only defined in the continuum, we formulate the above question in the continuum. To do this we take advantage of the variational approach developed in Section \ref{sec4}, in which we assigned the action $S[h]$ to a trajectory $h$ between two points in the plane. For the present problem, we use the same action (\ref{eq_non_deformed_action}) as when there was no constraint at all, and restrict it to those $h$ that do not cross the given curve; a justification for keeping the same action is provided in Section \ref{sec6}.

The following setting should be general enough. As we did before, we consider a trajectory in the plane as the graph of a continuous function $h(x)$. The general problem is, given a connected domain $\mathcal D$, whose boundary is sufficiently smooth (say, it has a tangent everywhere), to find the trajectories $h$ from $(0,0)$ to $(a,b)$, wholly contained in $\mathcal D$, which maximize the action  
\be
S[h] = \int_0^a \dif x \:L\big(h'(x)\big), \qquad h(0)=0,\, h(a)=b, \quad{\rm and} \quad  \big(x,h(x)\big) \in {\mathcal D} \textrm{ for all $x \in [0,a]$}.
\label{varprob}
\ee
We recall that $L$ is the Lagrangean function, strictly concave, associated to the directed walk under consideration, and computable in terms of the sets $S$ and $W$. Also the function $h$ must have its (left- and right-) derivatives in the domain of $L$, namely the interval $[t_{\rm min},t_{\rm max}]$, itself determined by the set $S$. This imposes certain constraints on $\mathcal D$ for the point $(a,b)$ to be reachable. The last two conditions are implicit in the rest of this section.

Our main result is the following 

\vspace{5truemm} \noindent
{\bf Theorem} {\it For the variational problem formulated above (connected and simply-connected domain $\mathcal D$, boundary $\partial {\mathcal D}$ of class ${\mathcal C}^1$), the action $S[h]$ given in (\ref{varprob}) with $L$ strictly concave, has a global maximum in the set of continuous piecewise ${\mathcal C}^1$ functions, attained for a unique function $h^*$ of class ${\mathcal C}^1$. The maximizing $h^*$ is universal in the sense that it does not depend on the specific form of $L$. The function $h^*$ is best described as the shortest path in $\mathcal D$ between the points $(0,0)$ and $(a,b)$: the graph of $h^*(x)$ has rectilinear pieces alternating with portions of the boundary of $\mathcal D$; moreover the linear parts of $h^*$ meet or exit the boundary $\partial \mathcal D$ tangentially.}

\bigskip
The proof relies in an essential way on the strict concavity of the Lagrangean function and on the simple form of the action $S[h]$, which depends on $h$ through its derivative only. It basically is a consequence the following lemma.

\vspace{5truemm} \noindent
{\bf Lemma}  {\it Fix $(x_i,y_i)$ and $(x_f,y_f)$ two points in the plane and consider the set of continuous and piecewise ${\mathcal C}^1$ functions connecting these two points, so that $h(x_i)=y_i$ and $h(x_f)=y_f$ (the functions $h$ are not restricted to be contained in a certain domain). Then the action $S[h] = \int_{x_i}^{x_f} \dif x \:L\big(h'(x)\big)$ has a unique maximum for $h^*$ given by the straight line.}

\bigskip \noindent
Let $t^* = \tfrac{y_f-y_i}{x_f-x_i}$ be the slope of the straight line $h^*$. Assume first that $h$ is of class ${\mathcal C}^1$. In this case the derivative $h'(x)$ for $x \in [x_i,x_f]$ takes all values in a finite interval $[h'_{\rm min},h'_{\rm max}]$. Since the average value of $h'(x)$ over $[x_i,x_f]$ is equal to $t^*$, the interval $[h'_{\rm min},h'_{\rm max}]$ must contain $t^*$. The function $L$ being strictly concave, its graph lies below that of any of its tangents, in particular it lies below its tangent at $t^*$, namely $L(h') < L'(t^*) (h' - t^*) + L(t^*)$ for all $h' \neq t^*$. If the interval $[h'_{\rm min},h'_{\rm max}]$ is not reduced to the singleton $\{t^*\}$, that is, if $h$ is not the straight line $h^*$, then 
\be
S[h] = \int_{x_i}^{x_f} \dif x \:L\big(h'(x)\big) < L'(t^*) (y_f - y_i) + (x_f - x_i) \big(L(t^*) - t^* L'(t^*)\big) = (x_f - x_i) L(t^*) = S[h^*].
\ee

In case $h$ is only piecewise ${\mathcal C}^1$, the image of $[x_i,x_f]$ under $h'$ is the union of intervals (when $h$ is not linear) and singletons (when $h$ is linear). The inequality $L(h') < L'(t^*) (h' - t^*) + L(t^*)$ used above nonetheless remains valid for any value $h'$ in the domain of $L$, and leads to the same conclusion $S[h] < S[h^*]$. \cqfd

%

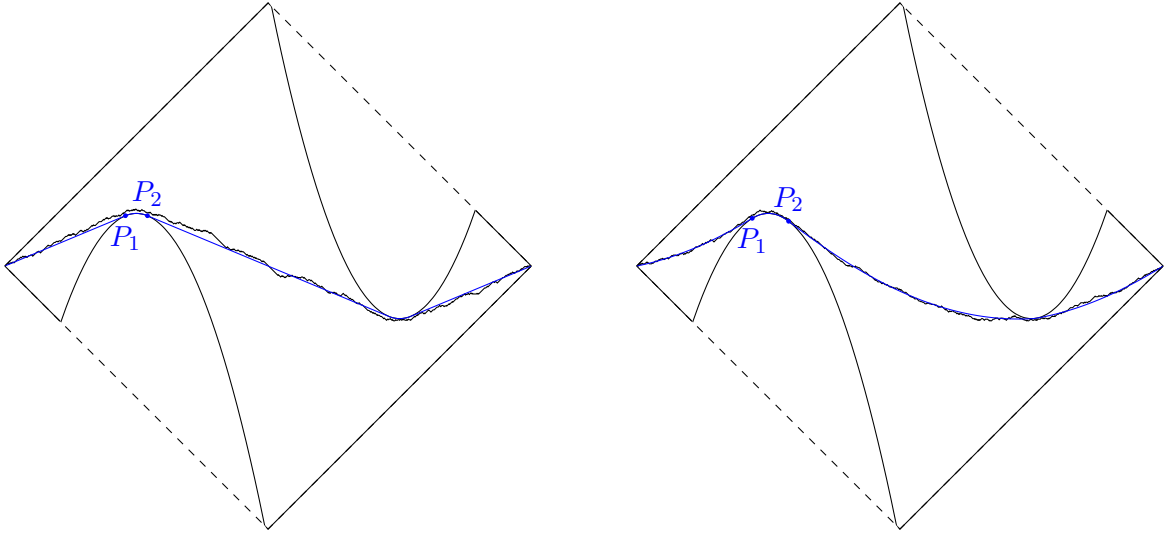
\begin{figure}[t]
\begin{center}
\begin{tikzpicture}[scale=0.7]
\begin{scope}
\draw (0,0)--++(5,5)--++(5,-5)--++(-5,-5)--++(-5,5);
\draw[color=white,fill=white,scale=10,domain=0.1:0.5,smooth,variable=\x] plot ({\x},{1/10 - 10*(\x - 1/4)*(\x - 1/4)});
\draw[color=white,fill=white,scale=10,domain=0.5:0.9,smooth,variable=\x] plot ({\x},{-1/10 +10*(\x - 3/4)*(\x - 3/4)});
\draw[dashed] (0,0)--++(5,5)--++(5,-5)--++(-5,-5)--++(-5,5);
\clip (0,0)--++(5,5)--++(5,-5)--++(-5,-5)--++(-5,5);
\draw[scale=10,domain=0.1:0.5,smooth,variable=\x] plot ({\x},{1/10 - 10*(\x - 1/4)*(\x - 1/4)});
\draw[scale=10,domain=0.5:0.9,smooth,variable=\x] plot ({\x},{-1/10 +10*(\x - 3/4)*(\x - 3/4)});
\draw [scale=1/9.6] plot file {example_shrodobs_N960_rescaled10.txt};
\draw[blue,scale=10,domain=0:sqrt(21)/20,smooth,variable=\x] plot ({\x},{(5-sqrt(21))*\x});
\draw[blue,scale=10,fill] ({sqrt(21)/20}, {1/10 - 10*(sqrt(21)/20 - 1/4)*(sqrt(21)/20 - 1/4)}) circle (0.1pt) node[below]{$P_1$};
\draw[blue,scale=10,domain=sqrt(21)/20:1/2-sqrt(21)/20,smooth,variable=\x] plot ({\x},{1/10 - 10*(\x - 1/4)*(\x - 1/4)});
\draw[blue,scale=10,fill] ({1/2-sqrt(21)/20}, {1/10 - 10*(1/2-sqrt(21)/20 - 1/4)*(1/2-sqrt(21)/20 - 1/4)}) circle (0.1pt) node[above]{$P_2$};
\draw[blue,scale=10,domain=1/2-sqrt(21)/20:1/2+sqrt(21)/20,smooth,variable=\x] plot ({\x},{-(5-sqrt(21))*(\x-1/2)});
\draw[blue,scale=10,domain=1/2+sqrt(21)/20:1-sqrt(21)/20,smooth,variable=\x] plot ({\x},{-1/10 +10*(\x - 3/4)*(\x - 3/4)});
\draw[blue,scale=10,domain=1-sqrt(21)/20:1,smooth,variable=\x] plot ({\x},{(5-sqrt(21))*(\x-1)});
\end{scope}
\begin{scope}[xshift=12cm]
\draw (0,0)--++(5,5)--++(5,-5)--++(-5,-5)--++(-5,5);
\draw[color=white,fill=white,scale=10,domain=0.1:0.5,smooth,variable=\x] plot ({\x},{1/10 - 10*(\x - 1/4)*(\x - 1/4)});
\draw[color=white,fill=white,scale=10,domain=0.5:0.9,smooth,variable=\x] plot ({\x},{-1/10 +10*(\x - 3/4)*(\x - 3/4)});
\draw[dashed] (0,0)--++(5,5)--++(5,-5)--++(-5,-5)--++(-5,5);
\clip (0,0)--++(5,5)--++(5,-5)--++(-5,-5)--++(-5,5);
\draw[scale=10,domain=0.1:0.5,smooth,variable=\x] plot ({\x},{1/10 - 10*(\x - 1/4)*(\x - 1/4)});
\draw[scale=10,domain=0.5:0.9,smooth,variable=\x] plot ({\x},{-1/10 +10*(\x - 3/4)*(\x - 3/4)});
\draw [scale=1/9.6] plot file {example_shrodobs_N960_q_rescaled10.txt};
\draw[blue,scale=10]  plot file {optimal_path_qdeformed.txt};
\draw[blue,scale=10,fill] ({0.219554}, {1/10 - 10*(0.2195540 - 1/4)*(0.219554 - 1/4)}) circle (0.1pt) node[below]{$P_1$};
\draw[blue,scale=10,fill] ({0.288598}, {1/10 - 10*(0.288598 - 1/4)*(0.288598 - 1/4)}) circle (0.1pt) node[above]{$P_2$};
\end{scope} 
\end{tikzpicture}
\end{center}
\caption{Numerical simulation of a single lattice path with elementary steps in $S=\{(1,1),(1,-1),(2,0)\}$ and a weight $w$ for the step $(2,0)$. The path starts from $(0,0)$ and ends at $(n,0)$ with $n=960$, and is constrained to stay in the domain delimited by the solid curve. On the left, the path is sampled with a weight $w=0.5$ and $q=1$, and on the right $w=1$ and $q=(0.05)^{1/n}$ (see Section \ref{sec7}).  The blue curves are the corresponding optimal trajectories $h^*(x)$, made of portions of the boundary and free trajectories (straight lines on the left, computed from \eqref{geodesic_ex2} on the right).}
\label{fig_zigzag_tangency}
\end{figure}

\vspace{5truemm}
To prove the Theorem, we take $(0,0)$ and $(a,b)$ to be the opposite vertices of a parallelogram (with slopes $t_\text{min},t_\text{max}$), in which the allowed region is delimited by the boundary $\partial {\mathcal D}$, see Figure \ref{fig_zigzag_tangency} (as we can see, the $\mathcal{C}^1$ property  is only required for certain portions of $\pa \mathcal{D}$). We repeatedly use the Lemma: to go from point $A$ to point $B$, the optimal path is the straight line, whenever this is possible (if the line does not cross the boundary). In particular, if $(a,b)$ is reachable from $(0,0)$ on a straight line, this will be the optimal $h^*$.

If not, the endpoint $(a,b)$, seen from the origin, is hidden by a portion of $\partial {\mathcal D}$. In this case, $h^*$ is a straight line starting from the origin that will necessarily make a first contact with $\partial {\mathcal D}$, say at $P_1$: it cannot go round $\partial {\mathcal D}$ without touching it because such a path could be optimized by following a shorter chord. Thus $h^*$ must be composed of straight lines and portions of $\pa \mathcal{D}$. $P_1$ must be the farthest point of the boundary that is reachable by a straight line from the origin hence must be a tangential point. Indeed, the Lemma implies that the union of a straight line to another contact point, and the portion of $\partial {\mathcal D}$ from it to the tangential contact point, has a smaller action. 

From $P_1$, the optimal path will follow $\partial {\mathcal D}$ for a while, exit the boundary at a certain point $P_2$ and follow a straight line towards some point $X$ (it could potentially follow the boundary till it reaches the endpoint, like for the Aztec diamond). $P_2$ is either a tangential point or else lies further down the boundary. Again the Lemma shows that a tangential point is optimal. Iterating these arguments proves the claim. It should be noted that the successive exit and passage points $P_2,X,\ldots$ cannot be precisely determined without knowing the shape of the domain that the paths will encounter further ahead. These points can only be determined from the global shape of $\mathcal D$. The resulting global process leads to the optimal path as depicted in Figure \ref{fig_zigzag_tangency}. The simply-connected assumption, usually satisfied in arctic curve situations, prevents the possibility to have two or more equally maximizing paths, ensuring the unicity of $h^*$.\cqfd

\medskip
In the examples of arctic curves known to the authors, the above result applies directly where the boundary $\partial {\mathcal D}$ is made of the arctic curve $\mathcal C$ itself and the boundary of the domain in which the model is naturally defined.


\section{Discussion}
\label{sec6}

We have taken a definite stand on how to treat the paths in presence of an impenetrable boundary, by restricting the variational problem obtained without boundary to trajectories contained in the allowed domain. If the endpoint is not reachable by a straight line, the optimal path will inevitably approach and coincide with portions of the constraining boundary. Since our derivation of the variational principle ignores the presence of a boundary, our approach calls for a justification. We provide it in this section. 

An alternative route would be to take into account, in the discrete lattice formulation itself, the constraint that the paths be contained in the allowed region, by modifying the weights of the elementary steps so as to prevent the walk from entering the forbidden region. In that formulation, one would compute the number of acceptable paths and the asymptotic form of the passage probabilities in order to figure out on what sort of trajectories the distribution concentrates in the scaling limit. Though probably more natural and more satisfactory, this approach is likely to be impractical for general domains as those considered above in the continuum, as it would require to handle position-dependent weights $w_i(\vec x)$. A (admittedly very) simple example is nonetheless worked out below which shows that the two approaches give identical results in the scaling limit. 

The main reason which validates our approach is based on the following argument. If it is true that the optimal path coincides with portions of the boundary in the continuum limit, it is not so on the lattice. The lattice formulation of the problem considers random lattice paths from $(0,0)$ to $(na,nb)$, and asks for the distribution over these paths, for large $n$, equivalently the distribution of passage on any given vertical line. In regions where the constraint plays no role, we know from Section \ref{sec32} that the distribution on paths has its support essentially contained in a window of size $n^{1/2}$ around the straight line (the average path). When the constraint is effective enough to force the lattice paths to deviate from their ballistic, straight line behaviour, one does not expect that the average path (or the likeliest, it does not make much difference) remains in a close neighbourhood of the constraint, because the entropy away from the constraint is higher: the paths which stay away from the constraint are more numerous. The entropic bias increases with the distance to the constraint and keeps the average path away from it. The average path will however not be pushed away from the constraint beyond a characteristic length scale of order $n^{1/2}$ since this is the scale of the fluctuations when there is no constraint. It is therefore natural to expect that the average path will be localized at a distance of order $n^{1/2}$ from the constraint. In this case, the derivation we made in Section \ref{sec4} of the action associated with a macroscopic path remains valid if we partition the horizontal interval $[0,na]$ into small cells whose size $\frac{an}K$ is of order $n^{1/2}$. Over this horizontal distance, the vertical displacement $(h_{j+1}-h_j)n$ is of order $\frac{an}K t_j$, also proportional to $n^{1/2}$. It then follows that the path trajectory is partitioned into windows of linear size $n^{1/2}$, big enough to contain the main fluctuations, and small enough to stay clear of the constraint. Inside each window, the path statistics is the same as if there were no constraint. So on the lattice, the constraint has a repulsive effect that moves the windows away from itself, but the statistics inside the windows remains unchanged, suggesting that the Lagrangean function is the same whether or not there is a constraint (this will be confirmed in the explicit example below). In the continuum limit, namely after rescaling by a factor $1/n$, the repulsion scales away and the restriction that the paths do not enter the forbidden region is then all what remains of the constraint.

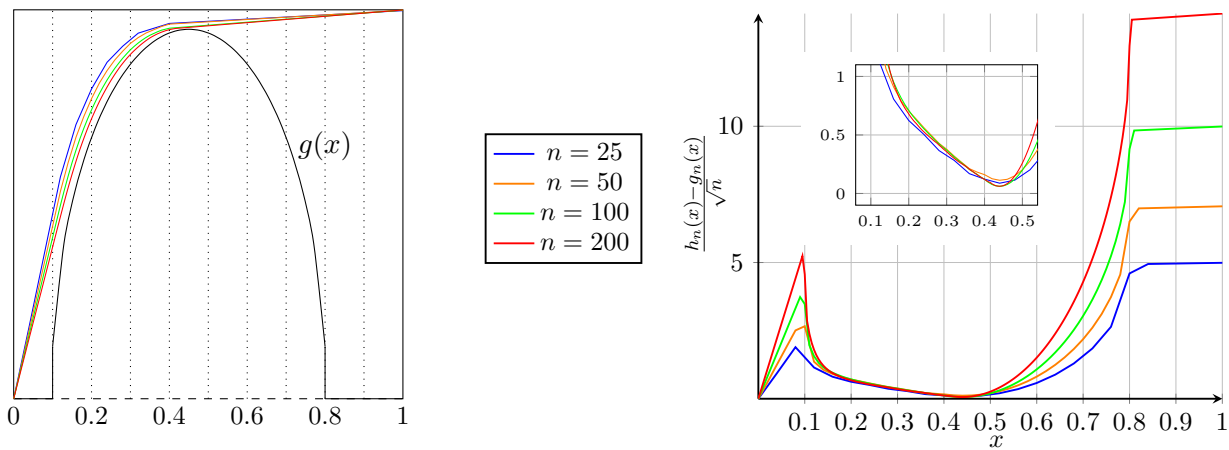
\begin{figure}[t]
\begin{center}
\begin{tikzpicture}[scale=0.9]
\begin{scope}[xshift=11cm]
\begin{axis}[
axis lines=middle
,xlabel={$x$}
,ylabel={$\frac{h_n(x)-g_n(x)}{\sqrt{n}}$}
,x label style = {at={(axis description cs: 0.55,-0.15)}}
,y label style = {at={(axis description cs: -0.05,0.35)},rotate=90,anchor=south west}
,xtick={0,0.1,0.2,0.3,0.4,0.5,0.6,0.7,0.8,0.9,1}
,grid
,thick
,legend style={at={(-0.25,0.35)},anchor=south east}
,xmin=0,xmax=1
]
\addplot[no marks,blue] table {rescaled_path_obs_N25.txt};
\addlegendentry{$n=25$}
\addplot[no marks,orange,mark repeat=2] table {rescaled_path_obs_N50.txt};
\addlegendentry{$n=50$}
\addplot[no marks,green,mark repeat=4] table {rescaled_path_obs_N100.txt};
\addlegendentry{$n=100$}
\addplot[no marks,red,mark repeat=8] table {rescaled_path_obs_N200.txt};
\addlegendentry{$n=200$}
\end{axis}

\node[fill=white] (zoom) at (2.5,3.7) {
    \begin{tikzpicture}
    \begin{axis}[
      tiny,
      ,xmin=0.1,xmax=0.5
      ,ymin=0,ymax=1
      ,enlargelimits
      ,ytick={0,0.5,1}
      ,grid
    ]
    \addplot[no marks,blue] table {rescaled_path_obs_N25.txt};
    \addplot[no marks,orange,mark repeat=2] table {rescaled_path_obs_N50.txt};
    \addplot[no marks,green,mark repeat=4] table {rescaled_path_obs_N100.txt};
    \addplot[no marks,red,mark repeat=8] table {rescaled_path_obs_N200.txt};
    \end{axis}
    \end{tikzpicture}
};
\end{scope}

\begin{scope}[scale=0.23]
\begin{scope}[scale=2.5]
\draw[scale=1,domain=1:8,smooth,variable=\x] plot ({\x},{10*0.95*sqrt((1-8*(\x-10*0.45)^2/100))});
\draw (1,0)--++(0,1.3435);
\draw (8,0)--++(0,1.3435);
\draw (8, 6.5) node{$g(x)$};
\draw[dashed] (0,0)--(10,0);
\draw (0,0)node[below,scale=0.9]{$0$}--(1,0);
\draw (8,0)--(10,0)node[below,scale=0.9]{$1$}--(10,10)--(0,10)--(0,0);
\foreach \i in {1,2,...,9}{
\draw[dotted] (\i,0)--++(0,10);}
\foreach \i in {2,4,6,8}{
\draw (\i,0)node[below,scale=0.9]{$0.\i$};}
\end{scope}
\draw[blue,scale=25]  plot file {rescaled_path_obs_N25_without.txt};
\draw[orange,scale=25]  plot file {rescaled_path_obs_N50_without.txt};
\draw[green,scale=25]  plot file {rescaled_path_obs_N100_without.txt};
\draw[red,scale=25]  plot file {rescaled_path_obs_N200_without.txt};

\end{scope}
\end{tikzpicture}
\end{center}
\caption{Results of numerical simulations for the walks with steps $(1,0)$ and $(0,1)$ as explained in the text. The paths shown on the left panel have been obtained by averaging over 100\,000 random paths for $n=25,50$ and over 40\,000 for $n=100,200$. The right panel shows the difference between the average path $h_n$ and the boundary $g_n(x)$ (given by $n$ times $g(x)$ drawn of the left panel) rescaled by a factor $\sqrt{n}$. The inset of the right panel focuses on the region where the paths stick to the boundary.}
\label{fig5}
\end{figure}

These expectations are borne out by the results of numerical simulations, shown in Fig. \ref{fig5}. The simulations were carried out for the walks made of the two elementary steps $(1,0)$ and $(0,1)$, each with weight 1 (the example 1 in Section \ref{sec2}), conditioned to start at $(0,0)$ and to end at the diagonal site $(n,n)$. The impenetrable boundary is represented by the parabola-like curve. The left panel shows the average paths, in the rescaled domain, for four values of $n$, whereas the right panel plots the distance between the average paths and the boundary $\partial {\cal D}$, rescaled by a factor $n^{1/2}$. In the region where the constraint is effective, one clearly sees that the average paths stay at all times at a distance of order $n^{1/2}$ from it.

Exact calculations are often possible when the constraint is piecewise linear. As a simple example, we consider the same walk as in the previous paragraph, with steps $(1,0)$ and $(0,1)$ each of weight 1. We condition the walk to start from the origin, to end at a distant site $(an,bn)$, with $\frac ba <1$, and we constrain it not to visit sites under the diagonal line $y=x$ for $0 \le x \le a_{\rm c} n < an$, in such a way that the endpoint $(an,bn)$ is not reachable along a single straight line. For $x > a_{\rm c} n$, the walk is no longer constrained although, due to the nature of the available steps, the walk will stay above (or on) the horizontal line $y = a_{\rm c} n$.

The lower boundary of the accessible domain $\cal D$ is made of two linear segments, a diagonal one and a horizontal one. We want to examine the alternative approach alluded to above, namely we solve the combinatorics of the constrained random paths, compute the continuous path $h^*$ on which the measure concentrates in the scaling limit and check that wherever the optimal path $h^*$ coincides with the diagonal constraint, the random lattice paths stay at a distance ${\cal O}(\sqrt{n})$ from it.

Let us denote by $Z_{r,s}$ the number of lattice paths from $(0,0)$ to $(r,s)$, and by $Z_{r,s}^{\rm c}$ the number of those which do not go under the diagonal line. They are explicitly given by
\be
Z_{r,s} = {r+s \choose r}, \quad (r,s \ge 0), \qquad Z_{r,s}^{\rm c} = \frac{s-r+1}{s+1} \, {r+s \choose r} \quad (s \ge r \ge 0).
\ee
Indeed they both satisfy the recurrence $a_{r,s} = a_{r-1,s} + a_{r,s-1}$ and the appropriate boundary conditions. These expressions make it clear that for $r,s \gg 1$ and fixed $t = \frac sr > 1$, $Z_{r,s}^{\rm c}$ is asymptotically equal to 
\be
Z_{r,s}^{\rm c} = \frac {t-1}t \sqrt{\frac{N_1(t)}{2\pi r}} \, \exp{\!\big\{r L_1(t)\big\}},
\ee
where the functions $N_1(t)$ and $L_1(t)$, given in Section \ref{sec2}, pertain to the walk without the constraint. As anticipated, the Lagrangean function is the same in the constrained and unconstrained cases.

Let us first look at the passage probability distribution on the vertical line $x=a_{\rm c} n$. Because vertical steps are allowed, we consider the probability ${\mathbb P}^{\rm 1st}_0[(a_{\rm c} n,a_{\rm c} n + m) | (an,bn)]$ that $(a_{\rm c} n, a_{\rm c} n + m)$ is the first visited site of the vertical line above the endpoint of the diagonal constraint, given that the path eventually reaches $(an,bn)$. This probability is proportional to the product of $Z^{\rm c}_{a_{\rm c} n-1, a_{\rm c} n+m}$ and $Z_{(a-a_{\rm c})n,(b-a_{\rm c})n-m}$. For large $n$, we find that the distribution is given by 
\be
{\mathbb P}^{\rm 1st}_0[(a_{\rm c} n, a_{\rm c} n + m) | (an,bn)] = N_\lambda \, \left(\frac m2+1 \right) e^{-\lambda m}, \qquad \lambda = \log{\Big[\frac 12 + \frac{a-a_{\rm c}}{2(b-a_{\rm c})}\Big]},
\ee
where $N_\lambda$ is the normalization factor. It follows that the average passage point remains at a distance ${\cal O}(1)$ above $(a_{\rm c} n, a_{\rm c} n)$, and implies that in the scaling limit, the continuous path $h^*$ visits the point $(a_{\rm c},a_{\rm c})$ with probability 1. Since after that point there is no effective constraint, $h^*$ is a straight line from $(a_{\rm c},a_{\rm c})$ to the final point $(a,b)$.

Let us now focus on the region above the diagonal constraint. As we have just seen that the passage distance to $(a_{\rm c} n,a_{\rm c} n)$ is of order ${\cal O}(1)$, we can consider the paths from $(0,0)$ to $(a_{\rm c} n,a_{\rm c} n)$ not going under the constraint. For $0 < \gamma < 1$, we look for the probability ${\mathbb P}^{\rm 1st}_0[(\gamma a_{\rm c}n,\gamma a_{\rm c}n+m) | (a_{\rm c}n,a_{\rm c}n)]$ that the first passage on the vertical line $x=\gamma a_{\rm c}n$ occurs at site $(\gamma a_{\rm c}n,\gamma a_{\rm c}n+m)$. We obtain that this probability is equal to 
\be
{\mathbb P}^{\rm 1st}_0[(\gamma a_{\rm c}n,\gamma a_{\rm c}n+m) | (a_{\rm c}n,a_{\rm c}n)] = \frac{Z_{\gamma a_{\rm c}n-1,\gamma a_{\rm c}n+m}^{\rm c} \times Z_{(1-\gamma )a_{\rm c}n-m,(1-\gamma )a_{\rm c}n}^{\rm c}} {Z_{a_{\rm c}n,a_{\rm c}n}^{\rm c}}.
\ee
For $n$ large and $m$ such that $\frac mn$ is small, we obtain the following asymptotic value,
\be
{\mathbb P}^{\rm 1st}_0[(\gamma a_{\rm c}n,\gamma a_{\rm c}n+m) | (a_{\rm c}n,a_{\rm c}n)] \simeq \frac{(m+1)(m+2)}{2\sqrt{\pi} [(\gamma (1-\gamma ) a_{\rm c}n]^{3/2}} \, \exp{\Big\{-\frac{m^2}{4\gamma (1-\gamma )a_{\rm c}n}\Big\}}, \qquad n \gg 1.
\ee
It implies that the average value $\langle m \rangle \simeq \frac{4}{\sqrt{\pi}} \sqrt{\gamma (1-\gamma)a_{\rm c}n}$ is of order $\sqrt{n}$ above the diagonal constraint.

In this particular example, we can conclude that indeed the random lattice paths stay at a distance ${\mathcal O}(\sqrt{n})$ away from the constraint, with the exception of a small neighbourhood of the extremal point where the two linear segments of the boundary meet, which is also the only point where the boundary is not ${\mathcal C}^1$. We can however note that the entropy bias advocated above becomes negligible for such a small region: the gain of entropy is only substantial when the paths keep away from the constraint over a macroscopic length scale. We expect these characteristics to be generic.

The above combinatorial analysis shows that in the scaling limit, the constrained random lattice paths condense with probability 1 on the single continuous path $h^*$ going from the origin to the point $(a_{\rm c},a_{\rm c})$ by following the diagonal constraint, and then goes straight from $(a_{\rm c},a_{\rm c})$ to $(a,b)$. This is also what the variational problem in the continuum yields. Despite the fact that this boundary is not ${\cal C}^1$, it is easy to see that the concavity argument implies that the optimal path $h^*$ in the (rescaled) domain is piecewise linear, and made of the two linear pieces mentioned above.


\section{Area weighted paths}
\label{sec7}

The tangent method is expected to be applicable in more general situations. A natural generalization of the previous discussion is a position-dependent weighting of the non-crossing paths for which the trajectory of a single path, without constraints, can differ from the straight line. In this case however, the assumption (i) of Section \ref{sec5} is expected to hold true. In this section we generalize the proof of the conjecture (ii) for a particular position-dependent weighting discussed in the context of the tangent method in \cite{DFG19,DFG19b}.

\subsection{Definition}

As before, we consider directed lattice paths with steps $S=\{\vec{s}_1=(u_1,v_1), \vec{s}_2=(u_2,v_2), \cdots, \vec{s}_k=(u_k,v_k)\}$ and weights $W=\{w_1,w_2,\cdots,w_k \}$. In addition, we assign each path a weight related to its area and controlled by a parameter $q>0$. For a path $p$ we define $A(p)$ to be the (signed) area between $p$ and the horizontal axis $y=0$. The total weight of $p$ is then the product of the weights of its elementary steps times  $q^{A(p)}$. This is a mild case of position dependent weights as it corresponds to assign a weight $w_i(x,y)=w_i \; q^{u_i(y + v_i/2)}$ to an elementary step $\vec{s}_i$ taken at $(x,y)$. 

We denote by $Z^q_{r,s}$ the weighted sum over all paths between $(0,0)$ and $(r,s)$. The generating function approach used to compute the numbers $Z_{r,s}$ can be extended to the $q$-weighted case. The main idea is to promote the variables $x$ and $y$ used in \eqref{gen} to non-commuting variables $\mathbf{x},\mathbf{y}$ satisfying $\mathbf{y}\mathbf{x}=q\mathbf{x}\mathbf{y}$. Defining $\mathbf{y}^{-1}$ by setting $\mathbf{y}^{-1} \mathbf{y}=\mathbf{y}\mathbf{y}^{-1}=1$, the previous relation then implies $\mathbf{y}^{-1}\mathbf{x}=q^{-1}\mathbf{x}\mathbf{y}^{-1}$. For a given set of elementary steps $S$ and weights $W$, we define 
\be
P^q(\mathbf{x},\mathbf{y}) = \sum_{i=1}^k w_i \, q^{u_i v_i/2} \; \mathbf{x}^{u_i} \mathbf{y}^{v_i}.
\ee
Here $w_i \, q^{u_i v_i/2}$ is the part of the weight $w_i(x,y)$ that is independent of the position $(x,y)$. The generating function can then be written
\be 
G^q(x,y) = \sum_{r,s}Z^q_{r,s} \; x^r y^s = \: :\!\frac{1}{1-P^q(\mathbf{x},\mathbf{y})}\!:
\label{qgen}
\ee
where $:\;\;:$ denotes the operation by which all the $\mathbf{x}$'s are moved to the left of the $\mathbf{y}$'s by using the above commutation relations, after what $\mathbf{x}$ and $\mathbf{y}$ are substituted by $x$ and $y$, respectively.  This reordering of the $\mathbf{x}$ and $\mathbf{y}$ variables brings in a power of $q$ which, combined with the partial weights contained in $P^q(\mathbf{x},\mathbf{y})$, exactly produces the right total weight of a path $p$, equal to $\big(\prod_i w_i) \, q^{A(p)}$. 

For an explicit example, let us reconsider the example 1 introduced in Section \ref{sec2}, namely $S=\{(1,0),(0,1)\}$ with $w_1=w_2=1$. In this simple case, $P^q(\mathbf{x},\mathbf{y})=\mathbf{x}+\mathbf{y}$ and we find
\be 
:\!\frac{1}{1-P^q(\mathbf{x},\mathbf{y})}\!: \: = \: :\! \sum_{n=0}^\infty (\mathbf{x}+\mathbf{y})^n \!: \: = \: :\!\sum_{n=0}^\infty \sum_{r=0}^n {n \choose r}_{\hspace{-0.15 cm}q} \, \mathbf{x}^r \mathbf{y}^{n-r}\! : \: = \sum_{r,s} {r+s \choose r}_{\hspace{-0.15 cm}q} \, x^r y^s,
\ee 
from which we deduce
\be
Z^q_{r,s}={
r+s \choose r }_{\hspace{-0.15 cm}q} = \prod_{k=1}^{r} \left( \frac{q^{k+s}-1}{q^k-1}\right).
\label{qpartitionfunction}
\ee
One readily checks that these numbers satisfy the recurrence relation $Z_{r,s}^q = q^s Z^q_{r-1,s} + Z^q_{r,s-1}$ and the appropriate boundary conditions.

\subsection{Variational approach}

For $q>1$ (resp. $q<1$), paths with an area close to the maximal (resp. minimal) value are more probable. However, in the scaling limit, trajectories with an extremal area are associated with less lattice paths and therefore have less entropy. Our first objective is to determine, in the continuum, the likeliest trajectory between two points, that is the trajectory which has the optimal balance between these two tendencies. We use the same notation as before for this curve, namely $h^*(x)$ (omitting the explicit dependence in $q$), and call it a free trajectory (or a geodesic) between $(0,0)$ and say $(a,b)$. 

In the discrete setting, when the endpoint $(r,s) = (na,nb)$ becomes large, the area of a generic path increases like $n^2$, so that $q$ must be rescaled in the scaling limit. As we will see, a non-trivial limit is obtained if we define $q=\mathfrak{q}^{1/n}$ for a fixed $\mathfrak{q}>0$; in the following we also set $\lambda=\log\mathfrak{q}$.

The variational approach of Section \ref{sec4} can be extended in a straightforward way. Consider again the collection of paths passing through the intermediate points $\vec{x}_j=(\frac{ja}{K} n,h_j n)$ for $j=0,\ldots,K$,  where $K$ is, as before, chosen to be proportional to $\sqrt{n}$. If $q$ were equal to $1$, their total weight would be equal to $Z_n[\{h_j\}]$ given in \eqref{eq_weighted_sum_q1} and associated with the weights of the elementary steps. However the paths $p$ contributing to $Z_n[\{h_j\}]$ do not all have the same area since $A(p)$ also depends on the intermediate passage points between the $h_j$'s. It turns out that the cumulated area of the rectangles of width $\tfrac{an}K$ and heights $n h_j$ provides a Riemann sum that is suitable to compute $A(p)$, the corrections to it bringing a contribution depreciated by a relative factor $1/\sqrt{n}$. Thus the dominant contribution of all these paths can be written in the form that only depends on the $h_j$'s,
\be 
Z^q_n[\{h_j\}]  \simeq Z^{}_n[\{h_j\}] \: q^{n^2 \sum_{j=0}^{K-1} \frac{a}{K} \, h_j}. 
\label{eq_area_qweight}
\ee
In the scaling limit, the first factor converges to $e^{n S[h]}$ for the action \eqref{eq_non_deformed_action} used earlier. The power of $q$ in the second term converges to $n^2 \int_0^a h(x)$ and justifies the scaling $q=\mathfrak{q}^{1/n}$ announced above, as the only way to obtain a non-trivial limit for large $n$: if the area term dominates, it leads to degenerate trajectories, and if it is dominated, we are back to the $q=1$ case. We obtain that the weight of a continuous trajectory $h$ has the following form,
\be
Z_n^q[h] \simeq e^{n S^q[h]}, \quad S^q[h]=\int_0^a \dif x \; L\big(h'(x)\big) + \lambda \int_0^a \dif x \;  h(x).
\label{eq_qweighted_action}
\ee
Note that one does not need to know $Z^q_{r,s}$ explicitly, since $L$ is computed solely from $S$ and $W$, as explained in Section \ref{sec2}. For Schr\"oder-like paths, the action has been computed in \cite{DFG19b} by using the explicit value of
$Z^q_{r,s}$ (a $q$-deformed trinomial coefficient) and exactly matches the form above upon
using the function $L_2(t)$ given in \eqref{ex2} (and an appropriate change of coordinates).

The problem now is to compute the free trajectories, namely  to maximize the functional \eqref{eq_qweighted_action} in the set of functions $h(x)$ that are continuous and piecewise $\mathcal{C}^1$, with 
\be
h(0)=0,\quad h(a)=b, \qquad t_{\text{min}} \le h'(x) \le t_{\text{max}}. 
\label{eq_conditions}
\ee
The strict concavity of $L$ ensures that the global maximum $h^*(x)$ is unique. Indeed, let us suppose that $h_1$ and $h_2$ are two distinct global maxima. Then $h_3=\frac{h_1+h_2}{2}$ (or any convex combination of $h_1$ and $h_2$) satisfies  \eqref{eq_conditions} and is such that $S^q[h_3]>S^q[h_1]=S^q[h_2]$, which is a contradiction. Using once more the strict concavity of the Lagrangean function $L$, the Weierstrass-Erdmann corner condition \cite{GF91} implies that $h^*(x)$ is in fact ${\cal C}^1$, and therefore satisfies the Euler-Lagrange equation over the entire domain,
\be
\frac{\dif}{\dif x} L'(h')  = \lambda, \qquad\qquad 0 \le x \le a.
\label{EL}
\ee

This equation can be explicitly solved. A first integration yields $L'(h'(x))=\lambda x + C_1$. Since $L''(t)<0$, $L'(t)$ is a strictly monotonic function and is therefore invertible. In fact, from Section \ref{sec2}, $L'(t) = -\log y(t)$ with $y$ itself being invertible ($y'(t)>0$), so we obtain
\be
h^*(x) = C_2 + \int_0^x \dif u \; y^{-1}\big(e^{-(\lambda u + C_1)}\big) = C_2 - \frac 1\lambda \int_{y^{-1}(e^{-C_1})}^{y^{-1}(e^{-(\lambda x + C_1)})} \dif v \: v \, \big(\!\log y(v)\big)',
\ee
where the second equality follows from the change of variable $e^{-(\lambda u + C_1)} = y(v)$. The relation $t \big(\log y(t)\big)' = -\big(\log x(t)\big)'$ obtained Section \ref{sec2} allows to carry out the remaining integration,
\be
h^*(x) = C_2 + \frac 1 \lambda \Big\{\log x\big(y^{-1}\big(e^{-(\lambda x + C_1)}\big)\big) - \log x\big(y^{-1}\big(e^{-C_1}\big)\big) \Big\},
\label{eq_general_hstar}
\ee
in terms of the functions $x(t)$ and $y(t)$ introduced in Section \ref{sec2}. The two integration constants are uniquely fixed by the boundary data, for instance the endpoints of the trajectory (\ref{eq_conditions}) (in which case $C_2=0$). On account of the relation $\log x\big(y^{-1}(e^{-z})\big) = z L'^{-1}(z) - L\big(L'^{-1}(z)\big)$, the solution $h^*(x)$ can be expressed solely in terms of $L$ (and the inverse of its first derivative). Whatever the form we give it, the unique solution $h^*$ is ${\cal C}^\infty$ in the variable $x$, and also in $C_1$ and $C_2$. 

For concreteness, we give the explicit form of the general solution for the two specific examples considered in Section \ref{sec2}. Using the functions $x(t), y(t)$ or $L(t)$ given there, we find (the constants $C_1,C_2$ have been redefined to $A,B$ in a $\lambda$-dependent way)
\bea
h_1^*(x) \egal \frac{1}{\lambda}\log(A-Be^{-\lambda x}), \qquad \lambda \neq 0,
\label{geodesic_ex1}\\
h_2^*(x) \egal x+B-\frac{1}{2 \lambda} \log\left(Ae^{2\lambda x}+(1+2w)+\sqrt{1+2(1+2w)Ae^{2\lambda x}+A^2 e^{4\lambda x} }\right) \nonumber\\
&&-\frac{1}{2 \lambda} \log\left((1+2w)Ae^{2\lambda x}+1+\sqrt{1+2(1+2w)Ae^{2\lambda x}+A^2 e^{4\lambda x} }\right), 
\qquad \lambda\neq 0.
\label{geodesic_ex2}
\eea
In each case, the geodesics appear as bent lines of constant concavity, as seen in Figure \ref{fig_zigzag_tangency}. Indeed the Euler-Lagrange equation $L''(h')\,h''=\lambda$ implies that they are strictly convex for $\lambda<0$, strictly concave for $\lambda >0$.

Contrary to the $q=1$ case, the free trajectories are not universal and depend on the elementary steps of the underlying walk as well as their weights. However, the tangency at the contact point between a free trajectory and a constraint is universal, as will be shown below. It is worth mentioning that the previous argument for the uniqueness of $h^*(x)$ is easily generalized for position dependent weights of the form $w_i(x,y)=w_i e^{u_i V(\frac{y}{n})}$ when $V$ is a smooth (not necessarily strictly) concave function. However for a general $V$, it is less clear how to construct a solution to the Euler-Lagrange equation, and one should rely on more abstract results to study the regularity of $h^*(x)$.

\subsection{Tangency}

We now turn to the case where the paths are constrained to stay inside a given domain $\mathcal{D}$. As before, we assume that the boundary $\partial\mathcal{D}$ is $\mathcal C^1$ in the regions where there is a contact between the optimal path $h^*$ and the boundary $\partial \mathcal D$ itself. The boundary repulsion mechanism discussed in Section \ref{sec6} is similar in the $q$-weighted case, so we will use the corresponding action \eqref{eq_qweighted_action} in the constrained situation.  Let us denote by $h^*(x)$ the solution of this maximization problem, and first prove that $h^*(x)$ is tangent to the boundary at the point of first contact $P_1$, if there is one (see figure \ref{fig_zigzag_tangency}).

Let us denote by $\phi_{X,Y}$ the free trajectory between the points $X$ and $Y$. The function $(X,Y,x) \mapsto \phi_{X,Y}(x)$ is $\mathcal{C}^\infty$ in $X,Y$ and $x$, as follows from \eqref{eq_general_hstar}. The uniqueness of the extremal solution \eqref{eq_general_hstar} between two points $X$ and $Y$ ensures that $h^*(x)$, the optimal trajectory in presence of constraints, can only consist of free trajectories and portions of the boundary. Indeed, if a part of $h^*$ is not a portion of the boundary, then it must make the action stationary and therefore be of the form \eqref{eq_general_hstar}.

If the endpoint $(a,b)$ is reachable from $(0,0)$, then $h^*$ is the free trajectory $\phi_{(0,0),(a,b)}(x)$ between these two points. If it is not, $h^*$ starts from the origin and follows a free trajectory until it reaches the boundary at $P_1$. Without loss of generality, we assume that the free trajectory hits the boundary from above. For convenience, we also view the boundary around $P_1$ as the graph of a function $g(x)$. It is clear that $P_1=(x_{\max},g(x_{\max}))$ with 
\be
x_{\max}=\max \,\{x>0 \: : \; \phi_{(0,0),(x,g(x))}(u) > g(u) \:\text{ for } 0 \le u < x\}.
\label{xmax}
\ee
Indeed, if the first contact point is at $x>x_{\max}$, it is not reachable by a free trajectory that starts from the origin, and if $x<x_{\max}$, one can optimize the path. It remains to show the tangency, namely that $\phi'_{\max}(x_{\max})=g'(x_{\max})$ with $\phi_{\max}=\phi_{(0,0),(x_{\max},g(x_{\max}))}=\phi_{0,P_1}$. 
\begin{enumerate}
\item If $\phi_{\max}'(x_{\max})>g'(x_{\max})$, there exists $x<x_{\max}$ such that $\phi_{\max}(x)<g(x)$, which is impossible since $\phi_{\max}$ is a valid trajectory (that is, it stays above the boundary).
\item Assume instead that $\phi_{\max}'(x_{\max})<g'(x_{\max})$. The $\mathcal{C}^1$ property of the free trajectories  ensures the existence of $x_0>x_{\max}$ such that $\phi_{(0,0),(x_0,g(x_0))}'(x)<g'(x)$ for all $x_{\max}\le x\le x_0$. Upon integration on $[x,x_0]$, this implies $\phi_{(0,0),(x_0,g(x_0))}(x) > g(x)$ for $x_{\max} \le x < x_0$. In addition $\phi_{(0,0),(x_0,g(x_0))}$ must also be above $\phi_{\max}$ for $x \in [0,x_{\max}]$ because the two curves cannot cross (they would have two points in common, the crossing point and the origin, and would therefore coincide). Thus we obtain $\phi_{(0,0),(x_0,g(x_0))}(u) > g(u)$ for $0 \le u < x_0$ with $x_0 > x_{\max}$, contradicting the definition of $x_{\max}$ in (\ref{xmax}).
\end{enumerate}

Hence we must have  $\phi'_{\max}(x_{\max})=g'(x_{\max})$.  The argument for the tangency of the point of first escape $P_2$ (if there is one) is similar. As before, the tangency at $P_1$ and $P_2$ implies the tangency of the subsequent contact points, which completes the proof.

We note that if the domain is simply-connected, the proof of the uniqueness of $h^*$ in presence of constraints generalizes. Indeed, let $h_1$ and $h_2$ be two distinct global maxima ($h_1\neq h_2$) of the constrained problem. Then $h_3=\alpha h_1 +(1-\alpha)h_2$ lies between $h_1$ and $h_2$ for $0<\alpha<1$. Since the domain is simply-connected and since $h_1$ and $h_2$ belong to it, $h_3$ is in the domain as well. It is therefore a valid trajectory and satisfies $S[h_3]>\alpha S[h_1]+(1-\alpha)S[h_2]=S[h_1]=S[h_2]$, which contradicts the assumption.

\section*{Acknowledgments}
We thank Gilles Parez for his careful reading of the manuscript. B.D. acknowledges 
the financial support of the Fonds Sp\'eciaux de Recherche (FSR) of the Universit\'e catholique 
de Louvain. P.R. is a Senior Research Associate of FRS-FNRS (Belgian Fund for Scientific Research). This work was supported by the Fonds de la Recherche Scientifique\,-\,FNRS and the Fonds Wetenschappelijk 
Onderzoek\,-Vlaanderen (FWO) under EOS project no 30889451.

\end{document}